\documentclass[onecolumn,sort&compress,numbers]{els-mrw} 

\usepackage{amsmath,amssymb,amsfonts,amsthm,makeidx,graphicx}
\usepackage{txfonts}
\usepackage{helvet}
\usepackage{subcaption}
\usepackage{multirow} 

\begin{document}


\chapter{Heavy quarkonia and new hadrons with two heavy quarks}\label{chap1}

\author[1]{Yuping Guo}%
\author[2]{Chang-Zheng Yuan}%

\address[1]{\orgname{Fudan University}, \orgdiv{Institute of Modern Physics}, \orgaddress{220 Handan Road, Shanghai 200433, China}}
\address[2]{\orgname{Institute of High Energy Physics}, \orgaddress{19B Yuquan Road, Beijing 100049, China}}

\articletag{Chapter Article tagline: update of previous edition, reprint.}

\maketitle

\begin{abstract}[Abstract]
 We give a pedagogical introduction to heavy quarkonia---bound states of a heavy quark and its antiquark (e.g., charmonium $c\bar{c}$, bottomonium $b\bar{b}$)---as well as to the exotic hadrons containing two heavy quarks that have been discovered since 2003. The review covers the foundational discoveries ($J/\psi$ and $\Upsilon$), basic properties, spectroscopy interpreted via potential models, production mechanisms at colliders, and decay modes. A significant focus is placed on the so-called ``$XYZ$" states---particles like the $X(3872)$, $Y(4260/4230)$, $Z_c(3900)$, and $P_c$---whose properties defy conventional quark model expectations. These states, considered candidates for hybrids, multi-quark states, hadronic molecules, or hadroquarkonia, provide unprecedented probes of non-perturbative QCD and challenge our understanding of quark confinement and hadron formation. The chapter summarizes the current experimental landscape and highlights key open questions driving future research in hadron spectroscopy. 
\end{abstract}

\begin{keywords}
   Hadrons\sep Exotic Hadrons\sep Quarkonium\sep Spectroscopy\sep Particle Production\sep Particle Decay\sep $XYZ$ Particles
\end{keywords}


\begin{glossary}[Nomenclature]
	\begin{tabular}{@{}lp{34pc}@{}}
		ATLAS	& A Toroidal LHC Apparatus \\
            BaBar   & BaBar Experiment \\
            Belle   & Belle Experiment \\
		BESIII	& Beijing Spectrometer \\
		BNL		& Brookhaven National Laboratory \\
		CDF		& Collider Detector at Fermilab \\
		CMS		& Compact Muon Solenoid \\
		D\O\    & DZero Experiment \\
		LEP		& Large Electron-Positron Collider \\
		LHC		& Large Hadron Collider\\
		LHCb	& Large Hadron Collider Beauty Experiment \\
		PDG		& Particle Data Group \\
		SLAC	& Stanford Linear Accelerator Center \\
	\end{tabular}
\end{glossary}

\section*{Objectives}

\begin{itemize}

\item \textbf{Foundational concepts}: Readers will learn the historical significance and basic properties of heavy quarkonium systems (charmonium and bottomonium), including their role in validating the Standard Model through landmark discoveries like the $J/\psi$ and $\Upsilon$ particles.

\item \textbf{Spectroscopy tools}: Readers will learn how potential models (e.g., the Cornell potential) are used to predict quarkonium energy levels and how these predictions compare with experimental observations of charmonium and bottomonium states.

\item \textbf{Production and decay mechanisms}: Readers will examine how quarkonia and new hadrons are produced in colliders and fixed-target experiments ($e^+e^-$ annihilation, ISR, $B$-decays, $pp/p\bar{p}$ collision, and $p\bar{p}$ annihilation) and explore their decay modes (OZI-suppressed/allowed, electromagnetic/hadronic transitions), using these as precision tests of both perturbative and non-perturbative QCD.

\item \textbf{Exotic hadrons}: Readers will explore unconventional states, such as the $X(3872)$, $Y(4260/4230)$, $Z_c(3900)$, and $P_c$, learning how to identify their exotic signatures and evaluate theoretical interpretations including multi-quark states, hadronic molecules, hybrids and other related scenarios.

\item \textbf{Open challenges}: Readers will gain insight into unresolved questions in the field, such as the missing conventional quarkonium states, the impact of coupled-channel effects in resonance parameter determination, and the searches for partners of observed exotic hadrons. 

\end{itemize}

\section{Introduction}\label{intro}

Heavy quarkonia are bound states composed of a heavy quark and its corresponding antiquark, such as charm-anticharm ($c\bar{c}$, charmonium) and bottom-antibottom ($b\bar{b}$, bottomonium) states. 
The discovery of the first heavy quarkonium state, the $J/\psi$ meson, in 1974~\cite{E598:1974sol, SLAC-SP-017:1974ind}, marked a groundbreaking moment in particle physics, confirmed the existence of the charm quark and provided vital experimental information for the development of quantum chromodynamics (QCD) and the Standard Model (SM). Since then, heavy quarkonia have become essential tools for testing QCD, providing crucial insights into both the perturbative and nonperturbative aspects of the strong force.

In the high-energy (short-distance) regime, where the strong coupling constant $\alpha_s$ is small, heavy quarkonia offer a clean environment for validating perturbative QCD (pQCD). The success of pQCD in describing quarkonium production and decay has established it as a powerful framework for understanding short-distance strong interactions. 
In contrast, at low energies (long distance), where $\alpha_s$ is large and perturbative methods are no longer applicable, models are needed to connect quarks with the hadrons observed in experiments. 
In this context, heavy quarkonia play a critical role in probing nonperturbative QCD, especially in understanding confinement, the mechanism that prevents quarks from existing in isolation.

The spectroscopy of heavy quarkonia is a powerful tool for studying the quark-antiquark interaction potential, often described by the Cornell potential: a Coulomb-like term at short distance, complemented by a confining linear term at long distance. 
Given the large masses of the charm ($\approx 1.5~\mathrm{GeV}$) and bottom ($\approx4.5~\mathrm{GeV}$) quarks, heavy quarkonium systems can be treated as nonrelativistic, allowing their energy levels to be calculated using the Schr\"{o}dinger equation. These theoretical predictions can then be directly compared with experimental data. Observed heavy quarkonium states can thus be categorized by their quantum numbers and energies within this framework.

Since the discovery of the $X(3872)$ in 2003~\cite{Belle:2003nnu}, a growing number of new hadrons have been observed in the charmonium and bottomonium mass regions. These states--often referred to as charmonium-like or bottomonium-like states---exhibit properties inconsistent with expectations of conventional quarkonium.
They are believed to be exotic hadrons, possessing more complex internal structures than the simple quark-antiquark configurations of traditional mesons or the three-quark configurations of baryons. 
Their existence challenges the simple picture of the conventional quark model and presents intriguing puzzles that have driven intense theoretical and experimental research. The study of these exotic states holds the potential to uncover new laws and could lead to significant breakthroughs in our understanding of strong interactions. 

This chapter reviews the current state of knowledge on heavy quarkonia and new hadrons containing two heavy quarks. 
Mesons composed of two heavy quarks with different flavors---such as $B_c$ states---are also included, as they are often discussed alongside heavy quarkonia.
We begin by summarizing the discovery of the charmonium and bottomonium states, $J/\psi$ and $\Upsilon$. 
Next, we discuss the general properties used to identify heavy quarkonia, followed by an overview of the production mechanisms of heavy quarkonia in colliders and their decay modes. These are key observables for testing theoretical models. 
The final section addresses the growing list of new hadrons containing two heavy quarks in the charmonium and bottomonium mass regions. 
Although a definitive classification of these states within exotic hadron categories remains unsettled, a coherent spectroscopy is gradually emerging. 
 

\section{Discovery of heavy quarkonia}
\label{sec1}

In 1974, the $J/\psi$ meson was discovered independently by two experimental groups: one at Brookhaven National Laboratory (BNL), led by Samuel Ting~\cite{E598:1974sol}, and the other at the Stanford Linear Accelerator Center (SLAC), led by Burton Richter~\cite{SLAC-SP-017:1974ind}. 

\begin{figure}[htb]
	\centering
	\begin{subfigure}{0.4\textwidth}
		\includegraphics[width=1.8 in, height=2.2 in]{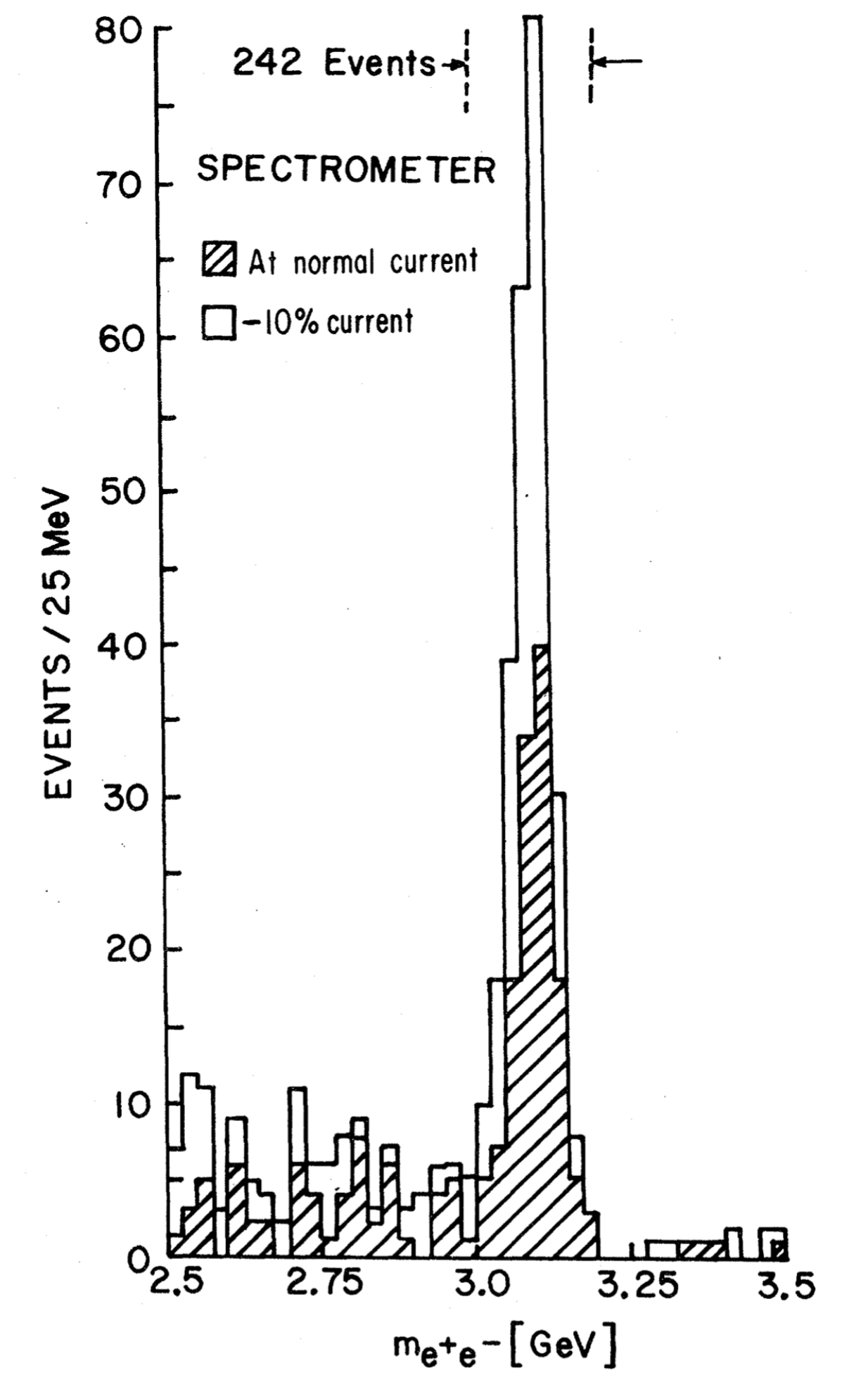}
		\caption{}
		\label{sec1:fig1:1}
	\end{subfigure}
	\begin{subfigure}{0.4\textwidth}
		\includegraphics[width=1.8 in, height=2.3 in]{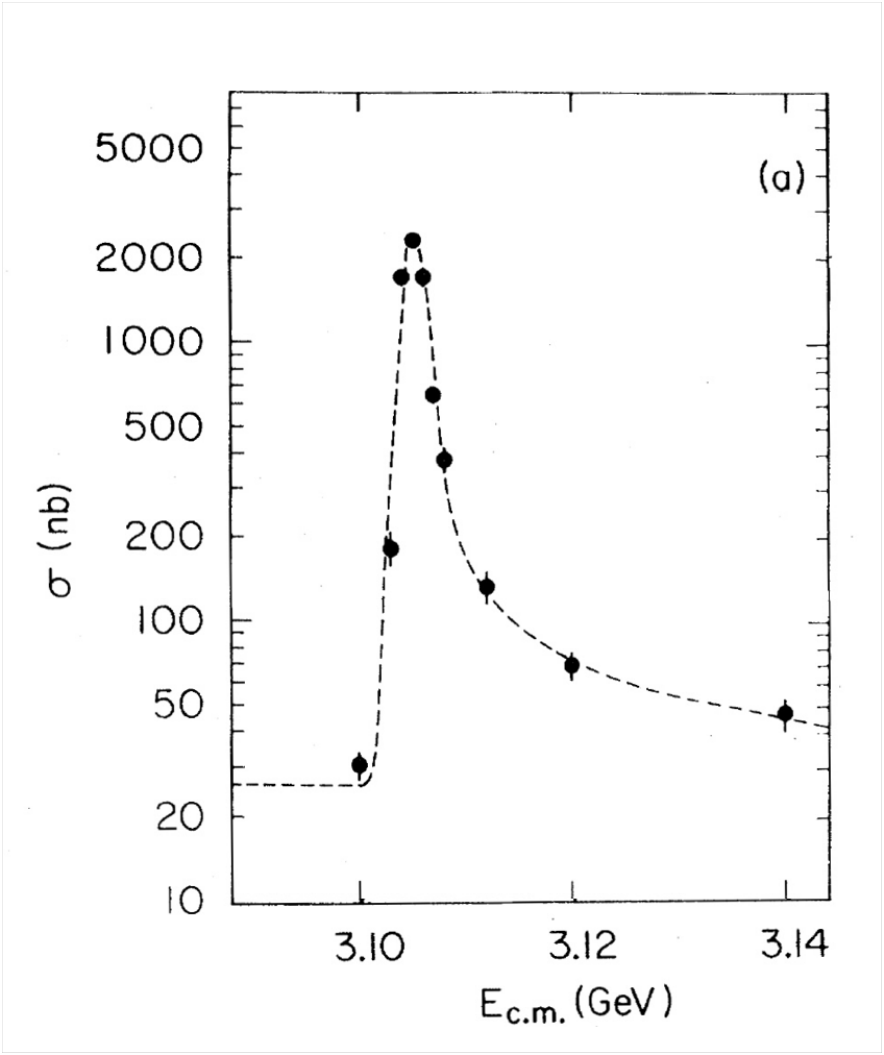}
		\caption{}
		\label{sec1:fig1:2}
	\end{subfigure}
	\caption{(a) The $J$ particle observed at BNL and (b) the $\psi$ particle observed at SLAC.}
	\label{sec1:fig1}
\end{figure}

At BNL, the observation was made by using a fixed-target experiment in which a $28.5~\mathrm{GeV}$ proton beam was directed onto a beryllium target. The reaction studied was
\begin{align}\label{sec1:eq1}
	p + \mathrm{Be} \to J + X \to e^+ e^- + X,
\end{align}
where $X$ denotes all particles that are not detected.
In this experiment, the invariant mass spectrum of the $e^+e^-$ pair was measured using a precise magnetic pair spectrometer. 
The invariant mass of the $e^+e^-$ pair was calculated as
 \begin{align}
 	m_{e^+e^-}^2=2m_e^2 + 2(E_{e^+} E_{e^-} - \vec{p}_{e^+} \cdot \vec{p}_{e^-} \cos(\theta_{e^+} + \theta_{e^-})),
 \end{align}
where $m_e$, $E_{e^{\pm}}$, and $(\vec{p}_{e^{\pm}}, \theta_{e^{\pm}})$ are the mass, energies, and (momenta, angles relative to the beam direction) of the electron and positron, respectively. A narrow peak at $m_{e^+e^-}=3.1~\mathrm{GeV}$ was observed in the invariant mass distribution, as shown in  Fig.~\ref{sec1:fig1:1}. This new particle was named the $J$ particle and was found to have an extremely narrow width, consistent with a lifetime of approximately $10^{-20}$ seconds. 

At SLAC, the observation was made using the MARK-I detector in an electron-positron collider experiment. The reaction studied was
\begin{align}\label{sec1:eq2}
	e^+e^- \to \psi \to \mathrm{hadrons},~e^+e^-,~\mu^+\mu^-,
\end{align}
The cross sections of the three processes were measured as a function of the center-of-mass (c.m.) energy ($\sqrt{s}$). A scan of the cross section between $3.1$ and $3.2~\mathrm{GeV}$ revealed a sharp peak, as shown in Fig.~\ref{sec1:fig1:2}. The new particle was named $\psi$, and its mass and width were determined to be
\begin{align}\label{sec1:eq3}
	E=3.105\pm0.003~\mathrm{GeV},~\Gamma \leq 1.3~\mathrm{MeV}.
\end{align}

The two particles observed were found to be the same, and it is now known as the $J/\psi$ meson. This discovery confirms the existence of the fourth quark predicted by Sheldon Lee Glashow, John Iliopoulos, and Luciano Maiani (GIM) in 1970~\cite{Glashow:1970gm}, and validates the GIM mechanism, which is the solution to the problem arising in the weak interaction theory with one charged vector boson coupled to the Cabibbo currents. 

In 1977, the $\Upsilon$ particle, identified as a bound state of $b\bar{b}$, was discovered by the E288 experiment at Fermilab, led by Leon Lederman~\cite{E288:1977xhf}. This experiment utilized a $400~\mathrm{GeV}$ proton beam incident on a fixed target, and the discovery was made through the reaction
\begin{align}\label{sec1:eq4}
	p+ (\mathrm{Cu, Pt}) \to \Upsilon + X \to \mu^+\mu^- + X.
\end{align}
Using a double-arm magnetic spectrometer, a statistically significant enhancement was observed at $9.5~\mathrm{GeV}$ in the invariant mass spectrum of $\mu^+\mu^-$ pairs, appearing as a sharp peak above an exponentially falling continuum. This resonance was named $\Upsilon$. The observation provided direct experimental evidence for the existence of the bottom (beauty) quark, confirming the necessity of a third generation of quarks and significantly extending the framework of the SM. 

\section{Basics properties}\label{sec3}

\subsection{Mass and width}\label{sec3:subsec1}

\begin{figure}[htb]
	\centering
	\includegraphics[width=1.8 in, height=2.3 in, angle=-90]{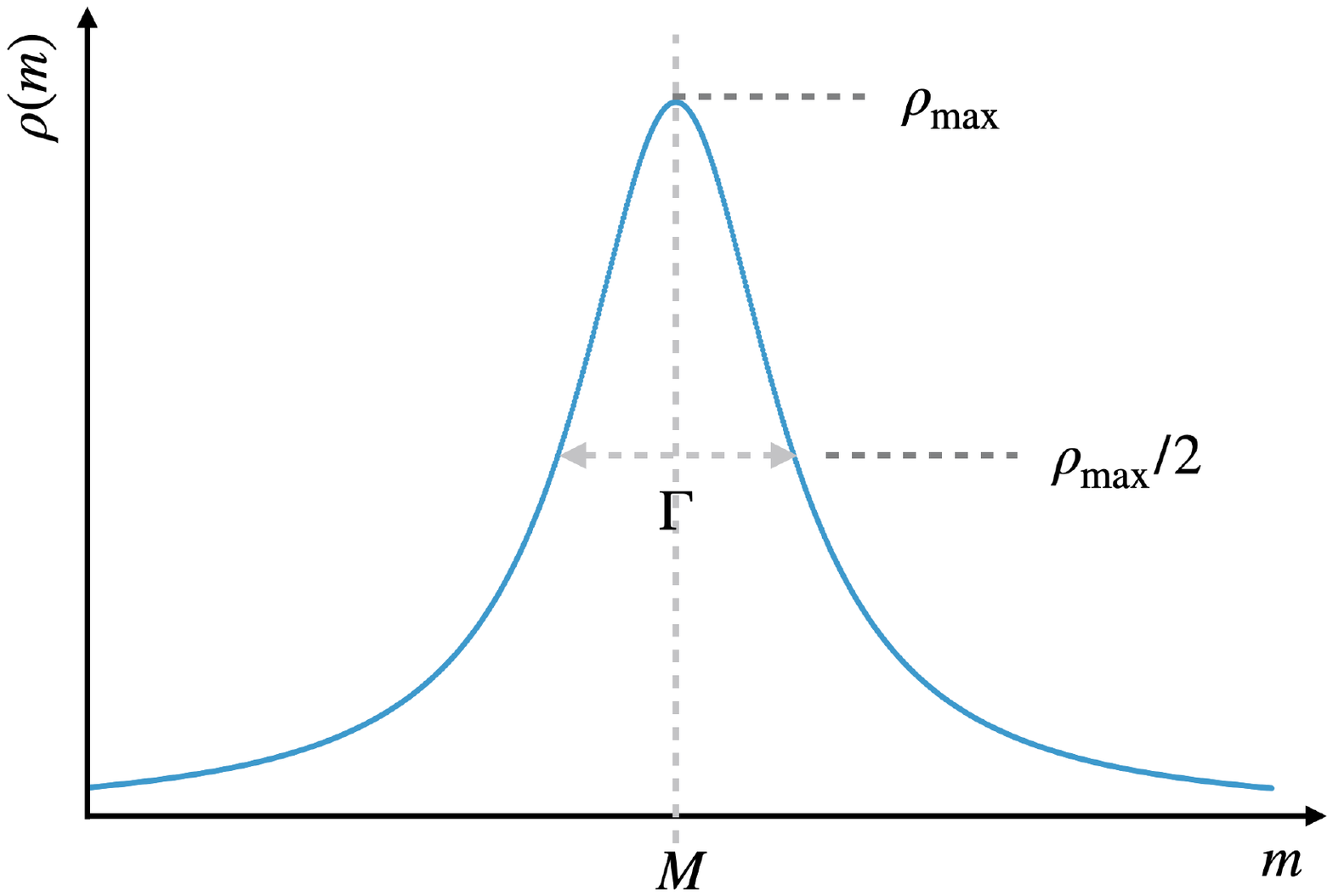}
	\caption{Plot of the Breit-Wigner formula.}
	\label{sec3:fig1}
\end{figure}

Heavy quarkonia are short-lived particles that decay primarily via strong interactions, are thus classified as resonances. The existence of these particles is inferred from the detection of the stable hadrons into which they decay. 
An isolated resonance away from strongly coupled thresholds is typically described by the Breit-Wigner formula
\begin{equation}\label{sec3:eq1}
   \rho(m)=\frac{K}{(m-M)^2+\Gamma^2/4},
\end{equation}
as illustrated in Fig.~\ref{sec3:fig1}. Here, $m$ represents the invariant mass of the final state particles, $M$ is the peak position of the distribution, $\Gamma$ denotes the full width of the peak at half maximum, and $K$ is a constant. 
The parameters $M$ and $\Gamma$ are used to characterize the mass and width (or decay width) of the resonance. The width $\Gamma$ is related to the mean lifetime of the state at rest, $\tau$, with the relation $\Gamma\equiv 1/\tau$.

\subsection{Branching fraction}\label{sec3:subsec2}

Heavy quarkonia can decay through multiple channels, and the probability of decay via a specific mode is described by the partial width $\Gamma_{i}$. The sum of all partial decay widths gives the width of the particle: $\Gamma=\sum_{i}\Gamma_i$. 
The branching fraction for a particular decay mode is defined as the ratio of the partial decay width for that mode to the total width: $\mathcal{B}_i = \Gamma_i/\Gamma$.

\subsection{Intrinsic property and notation}\label{sec3:subsec3}

Since the heavy quarkonium systems can be treated as nonrelativistic, there is a strong analogy to the electron-positron bound state, the positronium. The quarkonium states can be labelled by several quantum numbers: the orbital angular momentum $L$ between the quark and antiquark, the total spin $S$ of the system (with $S=0$ or $1$), the total angular momentum $\vec{J}=\vec{L}+\vec{S}$ (taking values between $|L-S|$ and $L+S$), and the radial quantum number $n$. These states are commonly denoted by the notation $n^{2S+1}L_{J}$ (sometimes simply noted as $nL$). 

The parity ($P$) and charge-parity ($C$) of the states are determined as 
\begin{align}\label{sec3:eq3}
	P=P_{Q}P_{\bar{Q}}(-1)^{L}=(-1)^{L+1}, ~~C=(-1)^{L+S},
\end{align}
where $P_{Q}$ and $P_{\bar{Q}}$ are the intrinsic parities of the heavy quark $Q$ and antiquark $\bar{Q}$, respectively. The allowed quantum numbers for heavy quarkonia are
\begin{align}\label{sec3:eq4}
    J^{PC}=0^{-+}, 1^{--}, 0^{++}, 1^{+-}, 1^{++}, 2^{++},  2^{--}, 3^{--}, \dots
\end{align}

\section{Spectroscopy}\label{sec4}

\subsection{Potential model}\label{sec4:subsec41}

In principle, the potential between a quark and an antiquark in a bound state can be derived directly from QCD; however, this is challenging in practice~\cite{QuarkoniumWorkingGroup:2004kpm}. Consequently, phenomenological models are often used to approximate the inter-quark potential. 
A commonly employed form is~\cite{PhysRevLett.34.369}
\begin{align}\label{sec4:eq1}
	V(r)=-\frac{4}{3}\frac{\alpha_s}{r} + k r,
\end{align} 
where $r=|\mathbf{r}|$ denotes the distance between the quark and antiquark, the factor $\frac{4}{3}$ arises from the SU(3) color factors, and $k$ is a parameter that can be adjusted to reproduce the experimental spectra. 
The first term corresponds to the short-distance Coulomb-like interaction resulting from one-gluon exchange, analogous to the one-photon exchange in quantum electromagnetic dynamics (QED). The second term accounts for the color confinement at long-distance through a linearly rising potential.

To achieve a more accurate description of the quarkonium spectrum, additional corrections are often included. The corrections include spin-dependent terms, such as the color contact term (spin-spin interaction), the color tensor interaction, and the spin-orbit interaction, as well as relativistic corrections and coupled-channel effects. Such refinements lead to fine and hyperfine structure splittings and are essential for precise calculations of energy levels, especially for higher-mass $Q\bar{Q}$ states above open-flavor mass thresholds, $m_{Q\bar{Q}}$ (the energy required to produce a $Q\bar{u}$-$\bar{Q}u$ meson pair). 
In turn, the mass splitting measured from experiments can be used to adjust or verify these additional corrections. 

\subsection{Charmonium and bottomonium spectroscopy}\label{sec4:subsec42}

\begin{figure}[htbp]
	\centering
	\begin{subfigure}{0.9\textwidth}
        \centering
		\includegraphics[width=\linewidth]{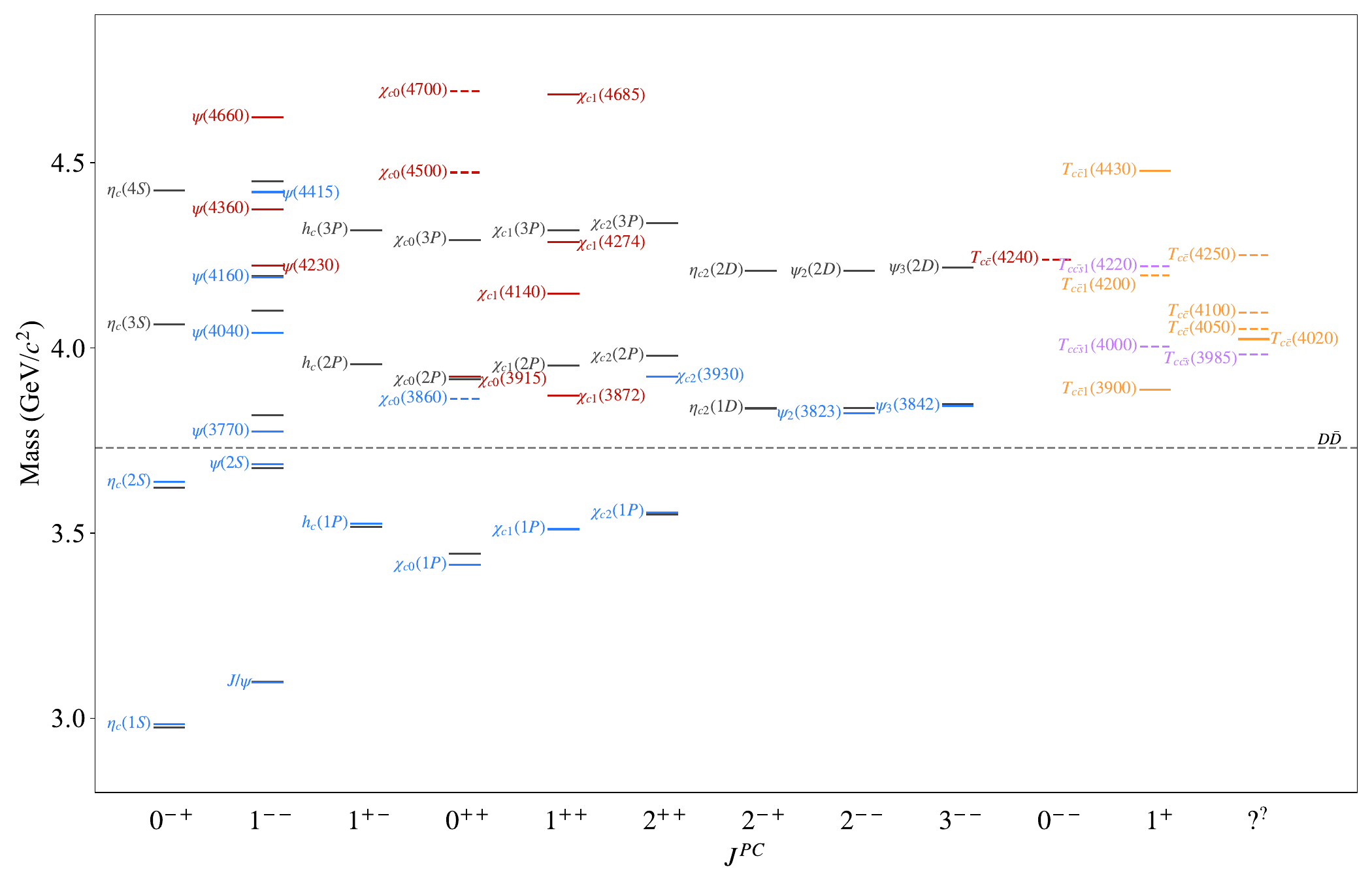}
		\caption{}
	\end{subfigure}
    \vspace{0.5em}
	\begin{subfigure}{0.9\textwidth}
        \centering
		\includegraphics[width=\linewidth]{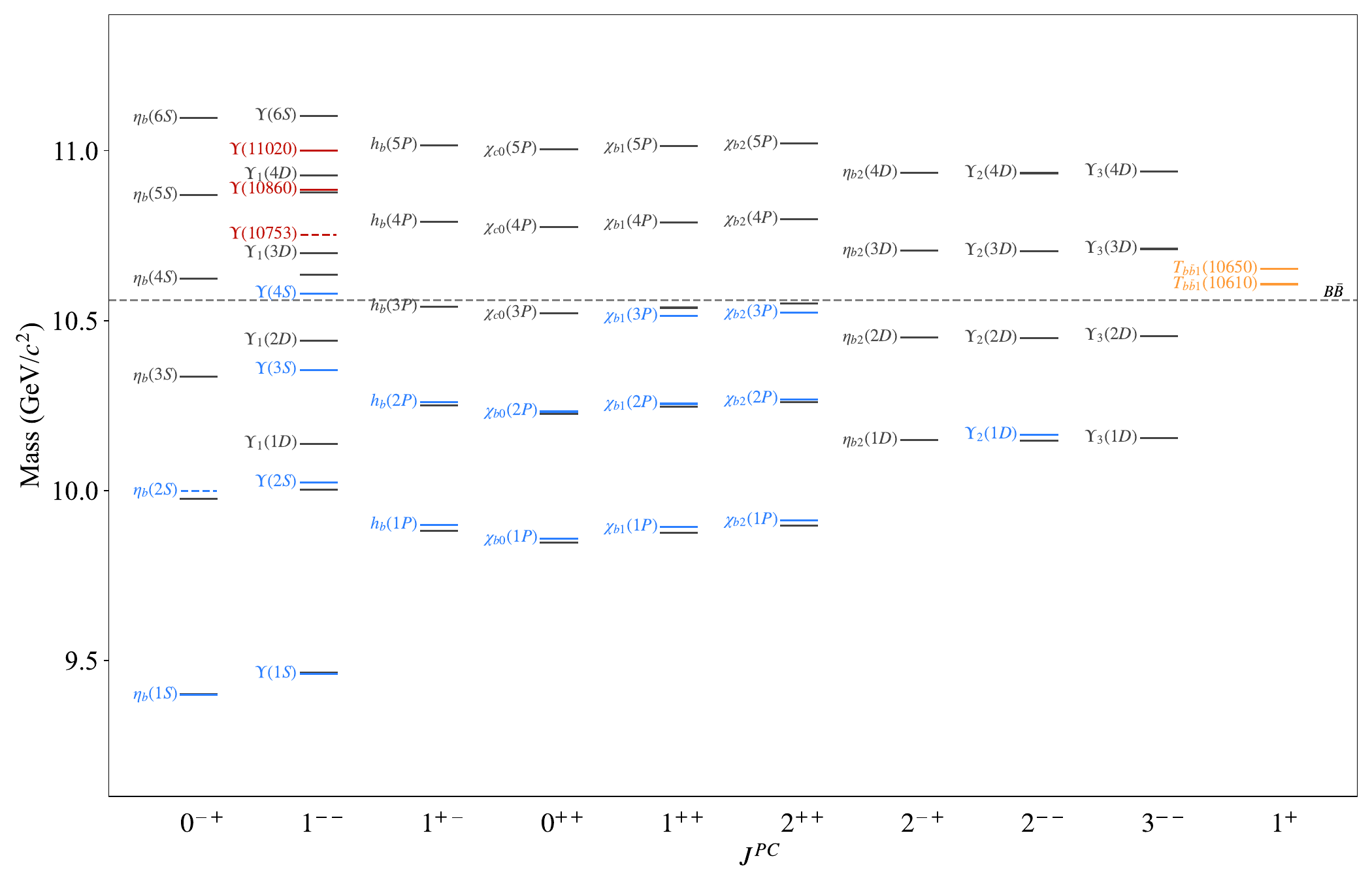}
		\caption{}
	\end{subfigure}
	\caption{The current status of the charmonium spectrum~\cite{Barnes:2005pb} (a) and the bottomonium spectrum~\cite{Godfrey:2015dia} (b).  Black solid lines indicate states predicted by potential model calculation, blue solid lines denote established states, blue dashed lines represent observed but not yet recognized as established states, and other colors correspond to exotic states. }
	\label{sec4:fig1}
\end{figure}

For a nonrelativistic quarkonium system, the dynamics in the c.m. frame of the $Q\bar{Q}=c\bar{c}$ or $b\bar{b}$ system can be described by the Schr\"{o}dinger equation
\begin{align}\label{sec4:eq2}
	-\frac{1}{2\mu}\nabla^2 \psi(\mathbf{r})+ V(r) \psi(\mathbf{r}) = E \psi(\mathbf{r}),
\end{align}
where $\mu$ is the reduced mass of the quark-antiquark system, $V(r)$ is the potential as described in Sec.~\ref{sec4:subsec41}, $\psi(\mathbf{r})$ is the spatial wave function, and $E$ is the interaction energy. 
The spectrum of charmonium and bottomonium states exhibits a structure analogous to that of the hydrogen atom. Predicted states from commonly referred calculations~\cite{Barnes:2005pb, Godfrey:2015dia} based on the Godfrey-Isgur relativized potential model~\cite{Godfrey:1985xj} are illustrated in Fig.~\ref{sec4:fig1} (black lines). 

The experimentally established states, as summarized by the Particle Data Group (PDG)~\cite{ParticleDataGroup:2024cfk}, are shown in Fig.~\ref{sec4:fig1} as solid blue lines. States observed in experiment but not yet recognized as established by the PDG are shown in dashed lines. 
In the charmonium system, all states below the lowest open-charm threshold ($m_{D\bar{D}}$) have been experimentally observed, and their measured properties generally agree well with predictions from potential models. However, above the open-charm threshold, the situation becomes more complicated: many predicted states remain unobserved, or their properties are not yet firmly established. 
A similar pattern is observed in the bottomonium system. In this case, more bound states lie below the open-bottom threshold ($m_{B\bar{B}}$), and some of these states have not yet been experimentally observed. 
Table~\ref{sec4:tab1} summarizes the properties of charmonium, $B_c$ (to be discussed in Sec.~\ref{sec4:subsec44}), and bottomonium states with masses below the lowest open-flavor mass thresholds. 

\begin{table}[htb]
	\TBL{\caption{Basic properties of $c\bar{c}$, $b\bar{c}$, and $b\bar{b}$ bound states below the lowest open-flavor thresholds, ordered by mass. Mass ($M$) and width ($\Gamma$) values are quoted from PDG 2024~\cite{ParticleDataGroup:2024cfk}. ``Production" refers to the production mechanism through which the state was first observed, and ``Decay" refers to the decay channel that exhibits the largest branching fraction measured to date. A question mark ($?$) indicates unmeasured quantities, while a dagger mark ($\dag$) denotes states that have not been considered established in the PDG. An $X$ represents undetected final-state particles.}
	\label{sec4:tab1}}
	{\begin{tabular*}{\textwidth}{@{\extracolsep{\fill}}@{}llllll@{}}
			\toprule
			\multicolumn{1}{@{}l}{\TCH{Name}} &
			\multicolumn{1}{l}{\TCH{M (MeV)}} &
			\multicolumn{1}{l}{\TCH{$\Gamma$ (MeV)}} &
			\multicolumn{1}{l}{\TCH{$I^{G}(J^{PC})$}} &
			\multicolumn{1}{l}{\TCH{Production}} &
			\multicolumn{1}{l}{\TCH{Decay}} \\
			\colrule
			$\eta_c(1S)$	& $2984.1\pm0.4$		& $30.5\pm0.5$			& $0^{+}(0^{-+})$	& $\psi(2S)\to \eta_c(1S)\gamma$ 	& $6\pi$ \\
			$J/\psi$		& $3096.900\pm0.006$	& $(92.6\pm1.7)\times 10^{-3}$	& $0^{-}(1^{--})$	& $pBe\to J/\psi X$				& $l^+l^-$\\
			                                          &					&				&		& $e^+e^-\to J/\psi$				&  \\
			$\chi_{c0}(1P)$	& $3414.71\pm0.30$ 	& $10.5\pm0.8$			&$0^{+}(0^{++})$& $\psi(2S)\to\chi_{c0}(1P)\gamma$ 	& $4\pi$ \\
			$\chi_{c1}(1P)$	& $3510.67\pm0.05$		& $0.84\pm0.04$			&$0^{+}(1^{++})$& $\psi(2S)\to\chi_{c1}(1P)\gamma$ 	& $ J/\psi\gamma$ \\
			$h_c(1P)$		& $3525.37\pm0.14$		& $0.78\pm0.28$			&$0^{-}(1^{+-})$ & $\psi(2S)\to h_c(1P)\pi^{0}$ 		& $\eta_c(1S)\gamma$ 		\\
			$\chi_{c2}(1P)$ & $3556.17\pm0.07$	& $1.97\pm0.09$ 			&$0^{+}(2^{++})$& $\psi(2S)\to\chi_{c2}(1P)\gamma$ & $J/\psi\gamma $ \\ 
			$\eta_{c}(2S)$	& $3637.8\pm0.6$		& $11.6\pm1.4$ 			& $0^{+}(0^{-+})$ & $B\to \eta_c(2S)K$			& $K\bar{K}\pi$ 			\\
			$\psi(2S)$		& $3686.097\pm0.010$	& $(286\pm16)\times 10^{-3}$	& $0^{-}(1^{--})$  & $e^+e^-\to\psi(2S)$			& $J/\psi \pi\pi$\\ 
			\midrule
			$B_c(1S)$	& $6274.47\pm0.32$ 	& $?$					& $0(0^{-})$ 	& $p\bar{p}\to B_c(1S) X$  			& $J/\psi \pi\pi\pi$\\ 
			$B_c(2S)$	& $6871.2\pm1.0$ 		& $?$ 					& $0(0^{-})$ 	& $pp\to B_c(2S) X$				& $B_c(1S)\pi^+\pi^-$ 		\\
			\midrule
			$\eta_b(1S)$ 	& $9398.7\pm2.0$		& $10^{+5}_{-4}$ 			 & $0^{+}(0^{-+})$& $\Upsilon(3S)\to\eta_b(1S)\gamma$ & $\mathrm{hadrons}$ 	\\
			$\Upsilon(1S)$ & $9460.40\pm0.10$ 	& $(54.02\pm1.25)\times 10^{-3}$ & $0^{-}(1^{--})$ & $p(Cu,Pt)\to \Upsilon(1S)X$ 	    & $l^+l^-$ \\
			$\chi_{b0}(1P)$ & $9859.44\pm0.42\pm0.31$	& $?$ 			 	& $0^{+}(0^{++})$  & $\Upsilon(2S)\to\chi_{b0}(1P)\gamma$ & $\Upsilon(1S)\gamma$\\
			$\chi_{b1}(1P)$ & $9892.78\pm0.26\pm0.31$ 	& $?$  				& $0^{+}(1^{++})$  & $\Upsilon(2S)\to\chi_{b1}(1P)\gamma$ & $\Upsilon(1S)\gamma$\\ 
			$h_{b}(1P)$ 	 & $9899.3\pm0.8$ 		& $?$					& $0^{-}(1^{+-})$   & $e^+e^-\to h_b(1P)\pi^+\pi^-$ 		& $\eta_b(1S)\gamma$ \\
			$\chi_{b2}(1P)$ & $9912.21\pm0.26\pm0.31$ & $?$ 					& $0^{+}(2^{++})$  & $\Upsilon(2S)\to\chi_{b2}(1P)\gamma$ & $\Upsilon(1S)\gamma$\\
			$\eta_b(2S)\dag$	 & $9999\pm4$			& $<24$					& $0^{+}(0^{-+})$	   & $\Upsilon(2S)\to\eta_b(2S)\gamma$	 & $\mathrm{hadrons}$	  \\ 
			$\Upsilon(2S)$  & $10023.4\pm0.5$ 		& $(31.98\pm2.63)\times10^{-3}$ & $0^{-}(1^{--})$  & $p(Cu,Pt)\to \Upsilon(2S)X$ 			& $\Upsilon(1S)\pi\pi$ \\
			$\Upsilon_2(1D)$ & $10163.7\pm1.4$	& $?$ 					& $0^{-}(2^{--})$	 & $\chi_{b1,2}(2P)\to\Upsilon_2(1D)\gamma$ & $\Upsilon(1S)\gamma\gamma$ \\
			$\chi_{b0}(2P)$ & $10232.5\pm0.4\pm0.5$ & $?$ 					& $0^{+}(0^{++})$ & $\Upsilon(3S)\to\chi_{b0}(2P)\gamma$ & $\Upsilon(2S)\gamma$ \\
			$\chi_{b1}(2P)$ & $10255.46\pm0.22\pm0.50$ & $?$ 				& $0^{+}(1^{++})$ & $\Upsilon(3S)\to\chi_{b1}(2P)\gamma$ & $\Upsilon(2S)\gamma$ \\ 
			$h_{b}(2P)$ 	& $10259.8\pm1.2$ 		& $?$ 					& $0^{-}(1^{+-})$  & $e^+e^-\to h_b(2P)\pi^+\pi^-$ 	&  $\eta_b(2S)\gamma$\\ 
			$\chi_{b2}(2P)$ & $10268.65\pm0.22\pm0.50$ & $?$				& $0^{+}(2^{++})$ & $\Upsilon(3S)\to\chi_{b2}(2P)\gamma$ & $\Upsilon(2S)\gamma$ \\
			$\Upsilon(3S)$  & $10355.1\pm0.5$ 		& $(20.32\pm1.85)\times10^{-3}$ & $0^{-}(1^{--})$& $p(Cu,Pt)\to \Upsilon(3S)X$			&  $\Upsilon(1S)\pi\pi$ 	\\
			$\chi_{b1}(3P)$ & $10513.4\pm0.7$		& $?$ 					&$0^{+}(1^{++})$ & $pp\to\chi_{b1}(3P)X$  		& $\Upsilon(1,2,3S)\gamma$\\
			$\chi_{b2}(3P)$ & $10524.0\pm0.8$ 		& $?$ 					&$0^{+}(2^{++})$ & $pp\to\chi_{b2}(3P)X$ 		& $\Upsilon(3S)\gamma$ 	\\
			\botrule
	\end{tabular*}}{%
	}%
\end{table}

Above the open-flavor threshold, only a limited number of states have been identified. In the charmonium sector, besides the $\psi(3770)$, three vector states---$\psi(4040)$, $\psi(4160)$, and $\psi(4415)$---have been established by fitting the inclusive hadronic cross section produced in $e^+e^-$ annihilation. These have long been considered as $3S$, $2D$, and $4S$ states. However, this assignment has been questioned by unquenched potential model calculations, which incorporate the effects of light quark-antiquark pairs in the vacuum. 
In bottomonium sector, the $\Upsilon(4S)$, $\Upsilon(10860)$, and $\Upsilon(11020)$ have been identified from the inclusive hadronic cross section. While the $\Upsilon(4S)$ is accepted as a conventional bottomonium state, the internal quark configurations of the latter two remain under debate.

\subsection{Toponium}\label{sec4:subsec43}

The top quark, carrying electric charge $+2/3$, is the partner of the bottom quark in the third generation of the SM. It is the heaviest known elementary particle, with a mass of approximately $173~\mathrm{GeV}$. 
Theoretical calculations based on nonrelativistic QCD and potential models predict the top-antitop bound state, toponium ($t\bar{t}$), decays rapidly after its formation, due to the extremely short lifetime of the top quark, about $0.5\times 10^{-24}$ seconds.

Unlike charmonium and bottomonium states, toponium would decay through the weak interaction of one of its constituent quarks, rather than via annihilation into gluons or photons. These states could manifest as threshold enhancements near the top-antitop production threshold together with angular observables characterising the spin and color connection of the $t\bar{t}$ pair, in both $e^+e^-$ collisions and hadron collisions. 

In 2025, the CMS Collaboration reported the observation of an enhancement near the $t\bar{t}$ threshold, consistent with the expected features of a ${}^1S_{0}$ toponium state~\cite{CMS:2025kzt}. 
Independent confirmation by other experiments, such as ATLAS, and improved theoretical modeling of top-quark threshold dynamics will be crucial for determining the true nature of this excess. 

\subsection{Doubly heavy quark-antiquark systems of different flavors}\label{sec4:subsec44}

There are three types of such configurations, $B_c^{-}$ ($b\bar{c}$), $T_b^{+}$ ($t\bar{b}$), and $T_c^{0}$ ($t\bar{c}$) and their antiparticles. The latter two should have a very short lifetime and a low production rate, and have never been studied experimentally. 

The $B_c$ mesons are so far the only observed doubly heavy quark-antiquark bound states composed of different heavy flavors. The ground state, $B_c(1S)$, with quantum numbers $J^{P}=0^{-}$, was first observed by the CDF Collaboration in $p\bar{p}$ collisions, through its semileptonic decay $B_c^{\pm}(1S)\to J/\psi l^{\pm}\nu$~\cite{CDF:1998ihx}, where $l$ denotes a charged lepton ($e$ or $\mu$). Subsequently, this meson has been confirmed by both the CDF and D\O\ Collaborations via two decay channels: $B_c^{\pm}(1S)\to J/\psi l^{\pm}\nu$ and $J/\psi \pi^{\pm}$. The lifetime of $B_c$ is: $\tau_{B_{c}}=(0.510\pm0.009)\times10^{-12}$ seconds, which is consistent with the general range of lifetimes for weakly decaying particles. 

Unlike charmonium and bottomonium states, the distinct flavors of the $b$ and $c$ quarks inside the meson prevent annihilation into gluons, excited $B_c$ states decay to lower-mass states via electromagnetic transition (photon emission) or hadronic transition (pion emission). The first radially excited state, $B_c(2S)$, was observed by the ATLAS Collaboration in its hadronic transition process, $B_c(2S)\to B_c(1S)\pi^+\pi^-$~\cite{ATLAS:2014lga}, while the orbitally excited $B_c(1P)$ states were observed by the LHCb Collaboration through radiative transitions~\cite{LHCb:2025uce}.

\section{Quarkonium production at colliders and quarkonium decay}\label{sec5}

\subsection{Production mechanisms}\label{sec5:subsec51}

Quarkonium states can be produced through various processes, each providing valuable insights into the dynamics of strong interactions. The primary production mechanisms include electron-positron ($e^+e^-$) collisions, hadron collisions ($pp$ or $p\bar{p}$), and fixed-target experiments. Table~\ref{sec4:tab2} lists typical colliders and fixed-target experiments contributing to the study of heavy quarkonia and new hadrons with two heavy quarks. Furthermore, lower-mass quarkonium states can be produced via radiative or hadronic transitions from high-mass quarkonium states. 

\begin{table}[htb]
	\TBL{\caption{Typical $e^+e^-$ colliders, $pp/p\bar{p}$ colliders, and fixed-target experiments contributing to the study of heavy quarkonia and new hadrons with two heavy quarks.}
	\label{sec4:tab2}}
	{\begin{tabular*}{\textwidth}{@{\extracolsep{\fill}}@{}llll@{}}
			\toprule
			\multicolumn{1}{@{}l}{\TCH{Collider}} &
			\multicolumn{1}{l}{\TCH{c.m. energy}} &
			\multicolumn{1}{l}{\TCH{Experiment}} &
			\multicolumn{1}{l}{\TCH{Production mechanism}} \\
			\colrule
			\multirow{5}{*}{$e^+e^-$}		& $6.9-10.9~\mathrm{GeV}$	& 	CLEO		& $e^+e^-$ annihilation, ISR, $\gamma\gamma$, double charmonium, $B$ decays	\\			
									& $3.67-4.26~\mathrm{GeV}$	&	CLEO-c		& $e^+e^-$ annihilation			\\
			\cmidrule{2-4}
									& \multirow{2}{*}{$9.5-11.2~\mathrm{GeV}$}	&	BaBar		& \multirow{2}{*}{$e^+e^-$ annihilation, ISR, $\gamma\gamma$, double charmonium, $B$ decays}					\\
									& 	&	Belle (II)		&		\\
			\cmidrule{2-4}
									& $1.8-5.6~\mathrm{GeV}$	&	BESIII		&	$e^+e^-$ annihilation, $\gamma\gamma$	\\
			\midrule
			\multirow{5}{*}{$pp/p\bar{p}$}	
									& \multirow{2}{*}{$1.96~\mathrm{TeV}$}	& CDF			&	\multirow{2}{*}{$p\bar{p}$, $b$-hadron decays} 		\\
									&								& D\O\			&				\\
			\cmidrule{2-4}
									& \multirow{3}{*}{$7-14~\mathrm{TeV}$}	& LHCb			&	\multirow{3}{*}{$pp$ , $b$-hadron decays}			\\
									&								& ATLAS		&				\\
									&								& CMS			&				\\
			\midrule
			\multirow{2}{*}{fixed target}	& \multirow{2}{*}{$2.9-3.7~\mathrm{GeV}$}				& E760   &	\multirow{2}{*}{$p\bar{p}$ annihilation} \\
			                               & 	            & E835	 &	\\
			\botrule
	\end{tabular*}}{%
	}%
\end{table}


\subsubsection{Electron-Positron Colliders}

In $e^+e^-$ colliders, the initial state is well-defined, providing a clean experimental environment for studying heavy quarkonium and probing underlying QCD effects. The c.m. energy of the machine can be precisely tuned, as demonstrated by the observation of the $J/\psi$ meson at SLAC. When the collider operates near the mass of the vector quarkonium states (with quantum numbers $J^{PC}=1^{--}$), these states can be directly produced through electron-positron annihilation via a virtual photon:
	\begin{align*} 
		e^+e^- \to \gamma^{*} \to Q\bar{Q}.
	\end{align*}
At higher c.m. energies, additional production mechanisms become accessible for lower-mass heavy quarkonium states. 

Taken charmonium as an example, the key mechanisms include:
 \begin{enumerate}[a.]
 
		\item Initial state radiation (ISR) process: 
		\begin{align*}
			e^+e^-\to \gamma_{\rm ISR}\gamma^* \to \gamma_{\rm ISR} c\bar{c}
		\end{align*}
  
		In an ISR process, a photon is emitted from the initial state, effectively reducing the c.m. energy to produce vector charmonium states. The states produced in this process share the quantum numbers of the virtual photon, $J^{PC}=1^{--}$. 
  
		\item $b$-hadron decays: 
		\begin{align*}
			b\bar{q} \to c\bar{c} + s\bar{q}, ~bqq \to c\bar{c} + s q q
		\end{align*}
  
		Charmonium states are produced via the weak decays of the $b$ quark within the hadron. In this case, the charmonium states are produced at a secondary vertex, a characteristic that can be used to distinguish this process from prompt production. This mechanism allows the production of charmonium states with all possible quantum numbers:
		\begin{align*}
			J^{PC}=0^{-+}, 1^{--}, 0^{++}, 1^{++}, 2^{++}, 1^{+-}, \dots
		\end{align*}
		
		\item  Two-photon fusion: 
		\begin{align*}
			e^+e^- \to e^+e^-\gamma^{(*)}\gamma^{(*)}\to e^+e^- c\bar{c}
		\end{align*}
  
       In this reaction, each initial-state particle emits a photon, and the two photons interact to produce a charmonium state. This process produces states with positive $C$-parity, such as $\eta_c$, $\chi_{c0}$, and $\chi_{c2}$. The allowed quantum numbers are $J^{PC}=0^{\pm+}$, $2^{\pm+}$, \ldots. If one of the two intermediate photons is virtual, then $1^{++}$ states can also be produced. 
		
		\item Two-photon interaction: 
		\begin{align*}
			e^+e^- \to \gamma^*\gamma^*\to c\bar{c}
		\end{align*}
  
		The allowed quantum numbers are the same as in the two-photon fusion process. This process is highly suppressed compared to the one involving a single virtual photon exchange. The first observation of such a process was made by the BESIII Collaboration in 2022 in the channel $e^+e^-\to \chi_{c1}(1P)$~\cite{BESIII:2022mtl}. Data samples should be accumulated at c.m. energies around the mass of the states under investigation. 
		
		\item Double charmonium production: 
		\begin{align*}
		e^+e^- \to \gamma^{*} \to c\bar{c} + c\bar{c}
		\end{align*}
  
		In this reaction, two charmonium states are produced simultaneously via a single virtual photon, with the two charmonium states having opposite $C$-parity. 

 \end{enumerate}
	
	These mechanisms also apply to bottomonium states production, although higher c.m. energies are required due to the larger mass of the bottom quark. Additionally, if the c.m. energy is sufficient to produce a $Z$ boson, as was the case at LEP~\cite{Straessner2010}, heavy quarkonium can also be produced via direct $Z$ boson decays, $Z\to Q\bar{Q} + X$. 

\subsubsection{Hadron Colliders}

In hadron colliders, quarkonium states can be produced through direct production or via decays of other hadrons, such as $b$-hadron. The latter has already been discussed in the context of electron-positron colliders. The former, in which the contribution from $b$-hadron decays is excluded, is often referred to as ``prompt production". 
	
Prompt production can occur through parton-fusion processes of the form
	\begin{align*} 
		ij\to Q\bar{Q}+k,
	\end{align*} 
where each $i$, $j$, and $k$ denotes a light quark, a light antiquark, or a gluon ($g$). Such processes occur at order $\alpha_s^3$. Higher-order contributions include the gluon-exchange process, the ``associated-production" process (where two $Q\bar{Q}$ pairs with the same flavor appear in the final state), and the gluon fragmentation process~\cite{Brambilla:2010cs}. In hadron colliders, quarkonium states with all possible quantum numbers can be directly produced. 	
 
\subsubsection{Fixed-target experiments}

In fixed-target experiments, heavy quarkonium states can be produced via photo-production ($\gamma + p \to Q\bar{Q} + X$), electro-production ($e + p \to e + Q\bar{Q} + X$), direct proton-antiproton annihilation, $p\bar{p}\to Q\bar{Q}$, and so on. E760 and E835~\cite{Garzoglio:2004kw} are typical proton-antiproton annihilation experiments dedicated to the charmonium state studies. Quarkonium states with all possible quantum numbers can be produced directly in such setups. 

\subsection{Decay modes}\label{sec5:subsec52}

The decays of quarkonium states also provide essential insights into the dynamics of the strong interaction. The primary decay modes include electromagnetic transitions, hadronic transitions, hadronic decays, radiative decays, leptonic decays, and rare or forbidden decays. Rare and forbidden decays occur via weak interaction or violate certain symmetries (i.e., $C$ parity or $P$ parity). These processes are often closely related to the search for new physics beyond the SM and will not be discussed further here. 

\begin{enumerate}%

\item Electromagnetic transitions:

Electromagnetic transitions between quarkonium states occur via the emission of a monochromatic photon:
	\begin{align*}
		(Q\bar{Q})_i \to \gamma + (Q\bar{Q})_f,
	\end{align*}
where $(Q\bar{Q})_i$ and $(Q\bar{Q})_f$ represent the initial and final quarkonium states, respectively. 
	
These transitions can be categorized into electric and magnetic transitions. Electric transitions do not alter the quark spin ($\Delta S=0$), and those that change the orbital angular momentum by one unit ($\Delta L=1$) are called electric dipole (denoted as $E1$) transitions. In contrast, magnetic transitions flip the quark spin ($\Delta S=1$), and those that preserve the orbital angular momentum ($\Delta L=0$) are called magnetic dipole (denoted as $M1$) transitions. The $E1$ transition rate is typically higher than that of the $M1$ transition. For example, the branching fractions $\mathcal{B}[\psi(2S)\to \gamma\chi_{c1}(1P)]=(9.75\pm0.27)\%$ ($E1$ transition) and $\mathcal{B}[\psi(2S)\to\gamma\eta_c(1S)]=(3.6\pm0.5)\times 10^{-3}$ ($M1$ transition).
	 
Electromagnetic transitions have served as important tools for the discovery of quarkonium states with quantum numbers other than $1^{--}$. The $\eta_c(1S)$ and $\chi_{cJ}(1P)$ ($J=0,1,2$) states were discovered through the radiative transition from $\psi(2S)$ states.
	
\item Hadronic transitions:

Hadronic transitions between quarkonium states involve the emission of light hadrons. The general form is:
	\begin{align*} 
		(Q\bar{Q})_i \to l.h. + (Q\bar{Q})_f,
	\end{align*}
where $l.h.$ stands for the emitted light hadron(s). Given that the masses of $(Q\bar{Q})_i$ and $(Q\bar{Q})_f$ range from a few hundred $\mathrm{MeV}$ to no more than two $\mathrm{GeV}$, the kinematically allowed $l.h.$ are predominantly one or two light hadrons, such as $\pi^{0}$, $\eta$, $\eta^\prime$, $\omega$, $\phi$, $\pi\pi$, or $K\bar{K}$. 
	
These transitions are important decay modes for heavy quarkonia, either serving as discovery channels of new states or as benchmarks for phenomenological model calculations. The first observed hadronic transition, $\psi(2S)\to\pi^+\pi^- J/\psi$, has a branching fraction of $(34.69\pm0.34)\%$---the largest among all $\psi(2S)$ decay modes~\cite{ParticleDataGroup:2024cfk}. The $P$-wave spin-singlet charmonium state, $h_c(1P)$, was discovered via the hadronic transition process $\psi(2S)\to\pi^0 h_c(1P)$,
while the bottomonium state $h_b(1P)$ was observed via the process $\Upsilon(10860)\to \pi^+\pi^- h_b(1P)$.
	
\item Hadronic decays:
	\begin{align*}
		(Q\bar{Q})_i \to \mathrm{gluons} \to l.h.
	\end{align*}
Quarkonium states below the open-flavor threshold decay through the annihilation of their constituent $Q$ and $\bar{Q}$ quarks. A typical Feynman diagram for such a process is illustrated in Fig.~\ref{sec5:fig1:1}. This kind of diagram can be split into two parts by cutting only the gluon lines. The decay rate of such a process is suppressed due to the Okubo-Zweig-Iizuka (OZI) rule~\cite{Okubo:1963fa, Zweig:1964ruk, 10.1143/PTPS.37.21}. These decay modes are referred to as OZI-suppressed processes. Note that the OZI rule applies only to the decays via the strong interaction.
	
Conversely, quarkonium states above the open-flavor threshold can decay via the Feynman diagram shown in Fig.~\ref{sec5:fig1:2}, which cannot be split merely by cutting the gluon lines. These decay modes are referred to as OZI-allowed processes. The final state of the decay mode contains a pair of open-flavor mesons. This is why the width of the $\psi(3770)$ state is approximately 100 times larger than that of the $\psi(2S)$.
The dominant decay mode of $\psi(3770)$ is $D\bar{D}$, with a branching fraction of $(93^{+8}_{-9})\%$~\cite{ParticleDataGroup:2024cfk}. For vector quarkonium states, the $Q$ and $\bar{Q}$ quarks can annihilate into a virtual photon, which then hadronizes into a hadronic final state. These are electromagnetic decays, where OZI suppression is not applicable.
	
	\begin{figure}[htb]
	\centering
	\begin{subfigure}{0.4\textwidth}
		\includegraphics[width=1.5 in, height=2.2 in, angle=-90 ]{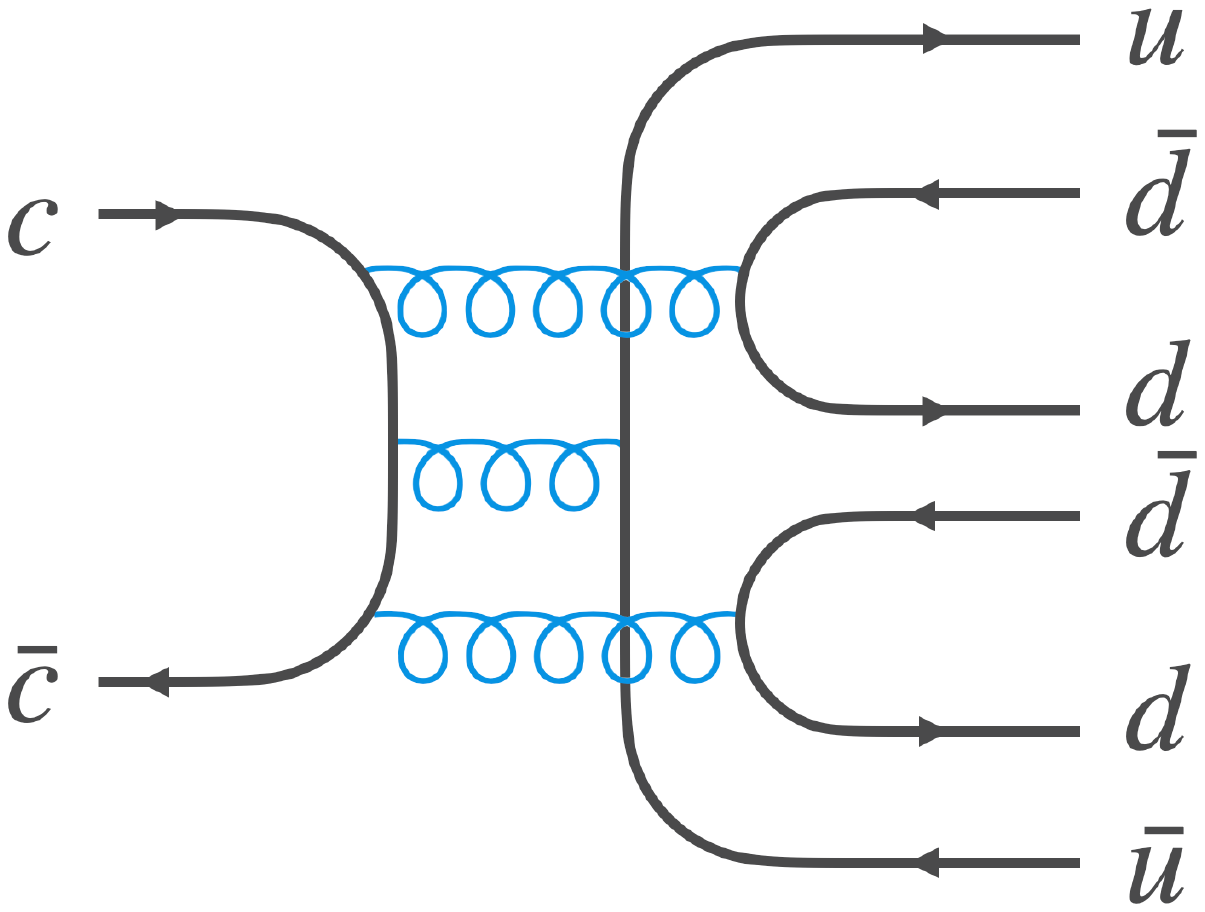}
		\caption{}
		\label{sec5:fig1:1}
	\end{subfigure}
	\begin{subfigure}{0.4\textwidth}
		\includegraphics[width=1.5 in, height=2.2 in, angle=-90]{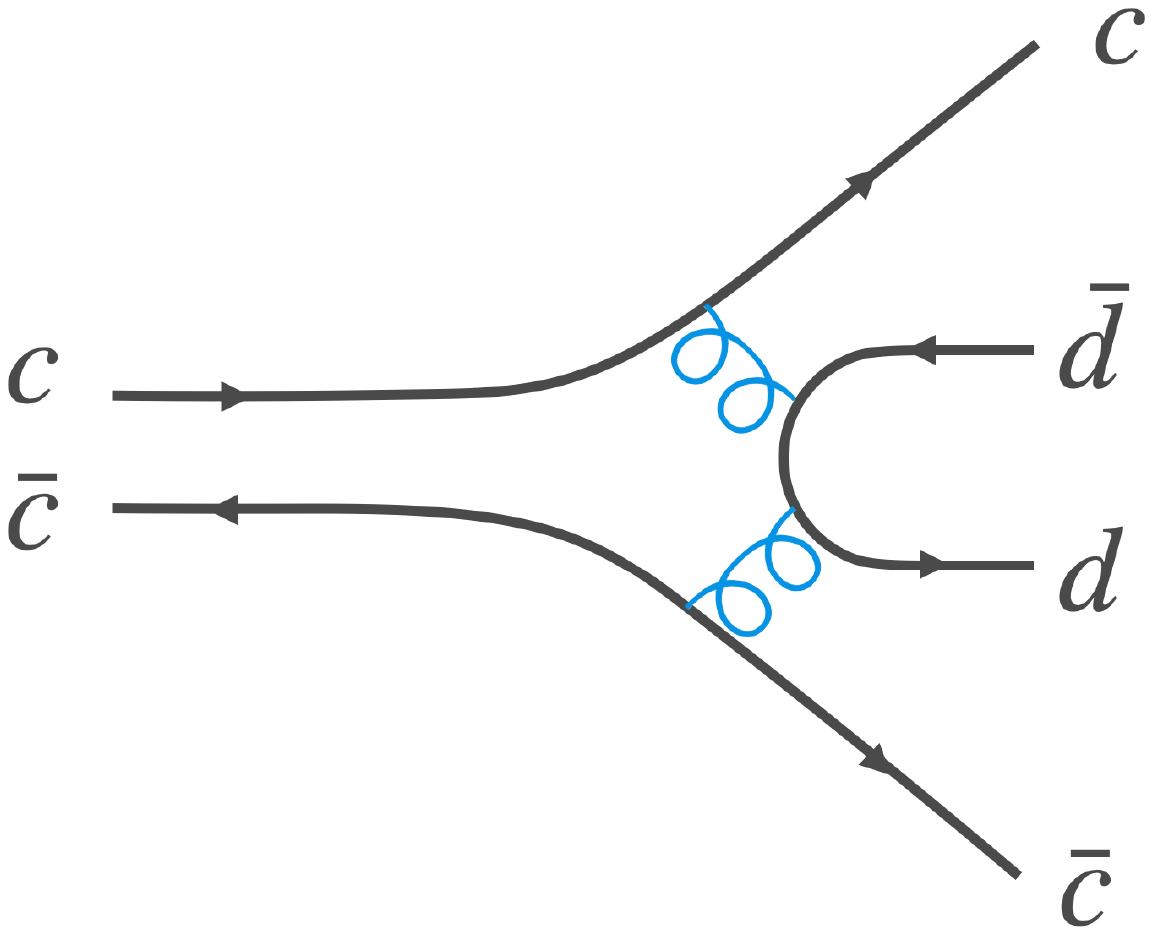}
		\caption{}
		\label{sec5:fig1:2}
	\end{subfigure}
	\caption{(a) OZI-suppressed decay mode and (b) OZI-allowed decay mode for charmonium state.}
	\label{sec5:fig1}
	\end{figure}
	
\item Radiative decays:

Radiative decays of quarkonium states proceed through a photon ($\gamma$) and at least two gluons:
	\begin{align*}
		(Q\bar{Q})_i\to \gamma + \mathrm{gluons} \to \gamma + l.h.
	\end{align*}
The gluons subsequently hadronized into light hadron(s). When the emitted photon is very soft (low-energy), this decay becomes indistinguishable from direct annihilation processes mediated by two or three gluons. Exclusive radiative decay rates are generally small; for example, the largest observed decay branching fraction for $J/\psi$ is $\mathcal{B}[J/\psi\to\gamma\pi^+\pi^-2\pi^0]=(8.3\pm3.1)\times 10^{-3}$. Radiative decay of quarkonium states remain poorly understood, both experimentally and theoretically. 
	
\item Leptonic decays:

The simplest parton-level decay of heavy vector quarkonium state occurs through annihilation of $Q$ and $\bar{Q}$ quarks into a virtual photon, which then decays into:
	\begin{enumerate}[a.]
		\item Dilepton pair: \\
		\begin{align*}
			e^{+}e^{-},~\mu^{+}\mu^{-}, ~\tau^{+}\tau^{-}
		\end{align*} 
These final state particles can be effectively reconstructed by modern detectors, enabling precise measurement of decay branching fractions. The leptonic width provides a measure of the wave function overlap at the origin, which is used to characterize fundamental features of the vector state. For scalar and tensor quarkonium states, analogous information is encoded in the two-photon decay width.   
		
\item Quark-antiquark pair: 

  \begin{align*}
			(Q\bar{Q})_i\to \gamma^* \to l.h. 
  \end{align*}
This reaction shares the same final state as hadronic decay via three gluons. However, this decay rate is proportional to the $R$ value: 
		$R\equiv\sigma[e^+e^-\to \mathrm{hadrons}]/\sigma[e^+e^-\to \mu^+\mu^-]$, 
where $\sigma$ represents the cross section.

\end{enumerate}
	
\end{enumerate}

\section{New hadrons with two heavy quarks}\label{sec6}

\subsection{Exotic hadrons}\label{sec6:subsec61}

The simple quark model, which proposed that mesons consist of a pair of quark and antiquark and baryons consist of three quarks, was introduced by M.~Gell-Mann~\cite{Gell-Mann:1964ewy} and G.~Zweig~\cite{Zweig:1964ruk} in 1964. This simple picture successfully categorized all known hadrons at that time and predicted the existence of the $\Omega$ baryon ($sss$), which was later confirmed experimentally. However, QCD allows for hadrons with more complex configurations, collectively referred to as exotic hadrons. These include (the quark flavors could be different):
\begin{enumerate}
   \item Glueballs: Hadrons composed solely of gluons  ($gg$, $ggg$, ...).
   \item Hybrids: Hadrons made of quarks with excited gluonic degrees of freedom ($q\bar{q}g$, $q\bar{q}gg$, ...).
   \item Multi-quark states: Hadrons with four or more quarks, such as tetraquark state ($q\bar{q}q\bar{q}$) and pentaquark state ($q\bar{q}qqq$).
   \item Hadroquarkonia: a heavy quark-antiquark core surrounded by a ``cloud" of light hadronic matter. 
   \item Hadronic molecules: loosely bound state of conventional hadrons.
\end{enumerate}

Tetraquarks, pentaquarks, hadroquarkonia, and hadronic molecules all contain more than three quarks, but their internal structure differs. The first two, also known as compact multiquarks, involve quarks tightly bound by direct interaction via color charge. In contrast, hadroquarkonia and hadronic molecules are composed of color-neutral clusters---traditional hadrons---that interact through the residual strong force, analogous to the deuteron as a bound state of a proton and a neutron. It should be noted that, in a broader sense, hadronic molecules are sometimes also classified as multi-quark states.


Despite decades of extensive searches, definitive evidence for exotic hadrons emerged only at the beginning of this century. Since the discovery of $X(3872)$ in 2003, numerous states that decay into a heavy quarkonium and light hadrons but with properties inconsistent with heavy quarkonium expectations have been discovered in the charmonium and bottomonium mass regions. These states are referred to as $XYZ$ states due to their uncertain nature. They are produced through the same mechanisms as quarkonium states described in Sect.~\ref{sec5:subsec51}. Moreover, exotic states with genuinely exotic quantum numbers---$J^{PC}$ combinations that cannot be formed by a simple quark-antiquark or three-quark configuration---can also be produced via those mechanisms, provided conservation laws are respected. A variety of theoretical models have been proposed to interpret the internal quark structure of these states; while some can explain certain observed states or some of their measured properties, none provides a universal explanation. 
For recent reviews, see Refs.~\cite{Esposito:2016noz, Guo:2017jvc, Ali:2017jda, Olsen:2017bmm, Brambilla:2019esw, Chen:2022asf}. 

It is also notable that some of the observed exotic hadron candidates are located close to the two-hadron mass thresholds. For example, $X(3872)$ is near the $D^{*0}\bar{D}^0$ threshold, and $Z_c(3900)^+$ is near the $D^{*+}\bar{D}^0$ threshold. Kinematic effects, such as cusps and triangle singularities, may play an important role in these observations~\cite{Guo:2019twa}. 

In the new naming scheme adopted by PDG in 2023, $XYZ$ states with identified quantum numbers are incorporated into the standard naming conventions. For example, the $X(3872)$ and $Y(4260)$ are now recognized as $\chi_{c1}(3872)$ and $\psi(4260)$, with the value in the parentheses representing the mass of the state. Symbols like $T_{\mathrm{quarks}}(\mathrm{mass})$ and $P_{\mathrm{quarks}}(\mathrm{mass})$ denote tetraquark and pentaquark states, respectively. These states were initially named $Z_c(\mathrm{mass})$ and $P_c(\mathrm{mass})$. In the subsequent sections, we will retain the $XYZ$ nomenclature for these new hadrons.

\subsection{$X$ states}\label{sec6:subsec62}

\subsubsection{$X(3872)$}\label{sec6:subsubsec621}

\begin{figure}[htb]
	\centering
	\includegraphics[width=2.0 in, height=2.5 in]{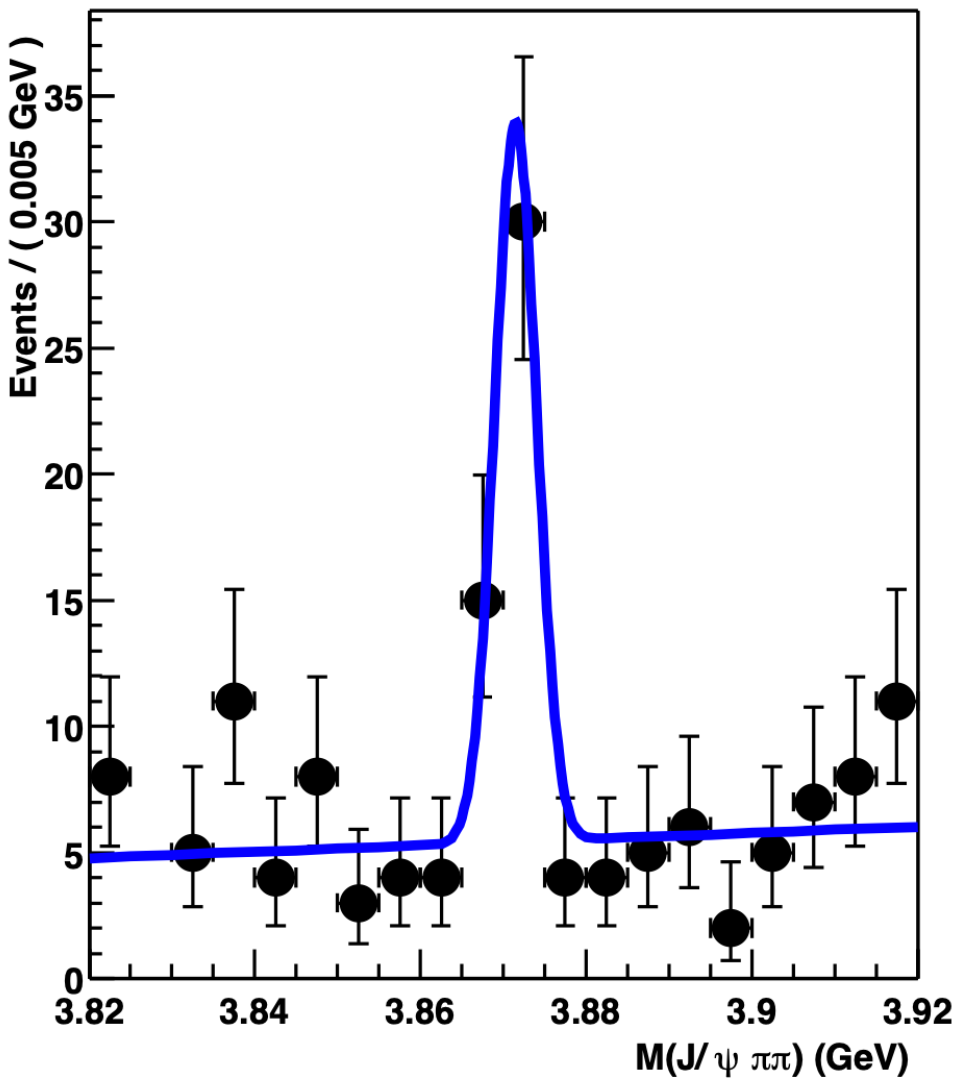}
	\caption{Invariant mass distribution of $J/\psi \pi^+\pi^-$ from $B$ meson decays~\cite{Belle:2003nnu}.}
	\label{sec6:fig1}
\end{figure}

The first $XYZ$ state, $X(3872)$, was discovered in 2003 by the Belle Collaboration in $B$ meson decays~\cite{Belle:2003nnu}:
\begin{align}
  B^{\pm}\to K^{\pm} + J/\psi\pi^+\pi^-.
\end{align} 
A distinct and narrow resonance peak was observed in the invariant mass spectrum of $J/\psi\pi^+\pi^-$ above the background events, as shown in Fig.~\ref{sec6:fig1}. 

This state was confirmed by several other experiments through different production mechanisms, including $pp/p\bar{p}$ collisions, $b$-hadron decays, radiative and hadronic transition from vector states, and two-photon fusion process. Although the properties of $X(3872)$ have been studied extensively, its internal quark composition remains unresolved. 

\begin{BoxTypeA}[sec5:box1]{Basic properties:}
	\begin{description}
	\item[Mass:] $3871.64\pm0.06$ MeV
	\item[Width:] $1.19\pm0.21$ MeV (Breit-Wigner width)  
	\item[Quantum numbers:] $I^{G}J^{PC}=0^{+}1^{++}$
	\item[Production:] $B$, $B_s$, and $\Lambda_b$ decays; $pp$, $p\bar{p}$, $p\mathrm{Pb}$, and $\mathrm{PbPb}$ prompt production; $e^+e^-\to X(3872)\gamma$ and $e^+e^-\to X(3872)\omega$; $e^+e^-\to e^+e^-\gamma\gamma^*\to e^+e^- X(3872)$
	\item[Decay:] (observed) $J/\psi\pi^+\pi^-$, $J/\psi\omega$, $\chi_{c1}(1P)\pi^0$, $J/\psi\gamma$, $\psi(2S)\gamma$, $D^{*0}\bar{D}^0$,  $D^0\bar{D}^0\pi^0$\\
	(not seen) $\eta_c(1S)\omega$, $\eta_{c}(1S)\pi\pi$, $D\bar{D}$, $\chi_{c1,2}(1P)\pi^{0}$, $\chi_{c0,1,2}(1P)\pi\pi$, $D\bar{D}\gamma$, $\chi_{c1,2}(2P)\gamma$, $\psi_2(3823)\gamma$, $p\bar{p}$, $\eta\pi\pi$
\end{description}
\end{BoxTypeA}

The quantum numbers of $X(3872)$ were determined by the LHCb Collaboration~\cite{LHCb:2013kgk} through a five-dimensional amplitude analysis of events from
\begin{align}
    B^+\to X(3872) K^+,~X(3872)\to J/\psi \pi^+\pi^-,~J/\psi\to\mu^+\mu^-.
\end{align}
Although $X(3872)$ shares the same quantum numbers as the $\chi_{c1}(2P)$ state, its mass is approximately $100~\mathrm{MeV}$ lower than predictions from potential models, as shown in Fig.~\ref{sec4:fig1}. Additionally, the ratio of the decay width of the isospin-violating process $X(3872)\to J/\psi\rho$ to the isospin-conserving process $X(3872)\to J/\psi \omega$:
\begin{align}
    \mathcal{B}[X(3872)\to J/\psi \rho]/\mathcal{B}[X(3872)\to J/\psi \omega]
\end{align}
is about orders of magnitude larger than that of a pure charmonium state. 
However, observed radiative decays to $J/\psi\gamma$ and $\psi(2S)\gamma$ suggest a sizeable charmonium component within $X(3872)$. 

The mass of $X(3872)$ lies very close to the $D^{*0}\bar{D}^0$ mass threshold, and the decay branching fraction of $X(3872)\to D^{*0}\bar{D}^0$ is about $50\%$~\cite{ParticleDataGroup:2024cfk}, which is the largest known decay mode. It is a natural candidate for a $D\bar{D}^{*}$ molecular state. However, the mass difference is only $(-0.05\pm0.12)~\mathrm{MeV}$, indicating a very loose binding. For comparison, the binding energy of deuteron is about $2.2~\mathrm{MeV}$. 
Both the hadronic molecular and tetraquark models predict partner states of $X(3872)$, but none have been observed so far. It is also possible that $X(3872)$ is an admixture of different components; then it is necessary to determine the fraction of each component.  

$X(3872)$ has been observed both in radiative transition and in hadronic transitions, accompanied by a $\gamma$ or an $\omega$ meson from $e^+e^-$ direct annihilation by the BESIII Collaboration~\cite{BESIII:2013fnz, BESIII:2022bse}. The radiative transition was observed using data sets collected near the $Y(4230)$---a candidate for a vector exotic hadron---while the hadronic transition was observed in data taken at $\sqrt{s}$ around $4.75~\mathrm{GeV}$, where $Y(4710)$ and $Y(4790)$ were reported. These findings suggest a possible connection between $X(3872)$ and the vector exotic charmonium states regarding their internal structure. 
 
\subsubsection{Other $X$ states near the predicted $\chi_{cJ}(2P)$ mass region and above}\label{sec6:subsubsec622}

Several states have been observed near the predicted $\chi_{cJ}(2P)$ mass region, including the $X(3860)$ ($\chi_{c0}(3860)$ in PDG), $X(3915)$ ($\chi_{c0}(3915)$ in PDG), and $\chi_{c2}(3930)$. These are considered candidates for the first excited $P$-wave triplet states, $\chi_{cJ}(2P)$. However, their assignment as $\chi_{c0}(2P)$ and $\chi_{c2}(2P)$ states remains inconclusive due to limited information on their production and decay properties. 

In addition, a rich spectrum of resonant structures above $4.0~\mathrm{GeV}$ has been observed in the $J/\psi\phi$ system, leading to the identification of five $X$ states with quantum numbers $J^{PC}=0^{++}$ or $1^{++}$. 

The basic properties of these $X$ states are summarized in Table~\ref{sec6:tab1}, while the states near the predicted $\chi_{cJ}(2P)$ mass region are discussed in more detail below. 

\begin{table}[htb]
	\TBL{\caption{Basic properties of $X$ states, ordered by mass.
	``Production" and ``Decay" refer to all the reported production mechanisms and decay modes; others are the same as in Table~\ref{sec4:tab1}.}
	\label{sec6:tab1}}
	{\begin{tabular*}{\textwidth}{@{\extracolsep{\fill}}@{}llllll@{}}
			\toprule
			\multicolumn{1}{@{}l}{\TCH{Name}} &
			\multicolumn{1}{l}{\TCH{M (MeV)}} &
			\multicolumn{1}{l}{\TCH{$\Gamma$ (MeV)}} &
			\multicolumn{1}{l}{\TCH{$I^{G}(J^{PC})$}} &
			\multicolumn{1}{l}{\TCH{Production}} &
			\multicolumn{1}{l}{\TCH{Decay}} \\
			\colrule
			$X(3860)\dag$		& $3862^{+50}_{-35}$	& $200^{+180}_{-110}$	& $0^{+}(0^{++})$	& $e^+e^-\to J/\psi X(3860)$ 	& $D\bar{D}$ \\
			$X(3915)$		& $3922.1\pm1.8$		& $20\pm 4$			& $0^{+}(0^{++})$	& $B\to X(3915) K$			& $D\bar{D}$ \\
						&					&					&				& $e^+e^-\to e^+e^-\gamma\gamma\to e^+ e^- X(3915)$	& $J/\psi \omega$ \\
						&					&					&				& $e^+e^-\to X(3915)\gamma$ 	& 	\\	
			$\chi_{c2}(3930)$	& $3922.5\pm1.0$	& $35.2\pm 2.2$		& $0^{+}(2^{++})$	& $B\to \chi_{c2}(3930) K$	&	$D\bar{D}$	\\
						&					&					&				& $pp\to \chi_{c2}(3930)X$	&	\\
						&					&					&				& $e^+e^-\to e^+e^-\gamma\gamma\to e^+ e^- \chi_{c2}(3930)$	&	\\	
			\midrule
			$X(4140)$		& $4146.5\pm3.0$		& $19^{+7}_{-5}$ 		& $0^{+}(1^{++}	)$	& $B\to X(4140) K$		 	& $J/\psi\phi$ \\
						&					&					&				& $pp\to X(4140)X$			&	\\
			$X(4274)$		& $4286^{+8}_{-9}$		& $51\pm 7$			& $0^{+}(1^{++})$	& $B\to X(4247) K$			& $J/\psi\phi$ \\
			$X(4500)\dag$	& $4474\pm4$			& $77^{+12}_{-10}$		& $0^{+}(0^{++})$	& $B\to X(4500) K$			& $J/\psi\phi$ \\
			$X(4685)\dag$	& $4684^{+15}_{-17}$	& $130\pm40$			& $0^{+}(1^{++})$	& $B\to X(4500) K$			& $J/\psi\phi$ \\
			$X(4700)\dag$	& $4694^{+16}_{-5}$		& $87^{+18}_{-10}$		& $0^{+}(0^{++})$	& $B\to X(4700) K$			& $J/\psi\phi$ \\

			\botrule
	\end{tabular*}}{%
	}%
\end{table}

\begin{itemize}

\item $X(3860)$:

The $X(3860)$ was observed by the Belle Collaboration in the double charmonium production process $e^+e^- \to J/\psi D\bar{D}$, appearing as a structure in the $D\bar{D}$ invariant mass spectrum~\cite{Belle:2017egg}. Based on the production mechanism, its $C$-parity should be positive. Although its quantum numbers have not been firmly established, the $0^{++}$ assignment is favored over $2^{++}$ by $2.5\sigma$ \footnote{In high energy physics, a signal with significance above $5\sigma$ is claimed as observation, above $3\sigma$ as evidence.}. Given its mass and decay mode into $D\bar{D}$, $X(3860)$ is suggested as a candidate for $\chi_{c0}(2P)$ state. However, both its existence and the $\chi_{c0}(2P)$ assignment require further confirmation. 

\item $X(3915)$:

The $X(3915)$ ($\chi_{c0}(3915)$ in PDG) is another intriguing state with $J^{PC}=0^{++}$. It was observed by the Belle Collaboration in the two-photon fusion process $e^+e^-\to e^+e^-\gamma\gamma\to e^+e^- X(3915)$, with $X(3915)$ decaying into $J/\psi \omega$~\cite{Belle:2009and}. This state was later confirmed by the BaBar Collaboration using the same decay channel. The $X(3915)$ is believed to be the same state reported earlier in the $B$ meson decays, $B\to J/\psi\omega K$, again first observed by Belle and subsequently confirmed by BaBar. An amplitude analysis of $B\to D^+D^- K$ events by the LHCb Collaboration provided the determination of its quantum numbers~\cite{LHCb:2020pxc}. Additionally, the BESIII collaboration reported a structure in this mass region through the radiative transition $e^+e^-\to \gamma\omega J/\psi$ in the $J/\psi\omega$ invariant mass spectrum.

\item $\chi_{c2}(3930)$:

A candidate for the $\chi_{c2}(2P)$ state, initially named as $Z(3940)$ ($\chi_{c2}(3930)$ in PDG), was observed by the Belle Collaboration and confirmed by the BaBar Collaboration in the two-photon fusion process $e^+e^-\to e^+e^-\gamma\gamma\to e^+e^-D\bar{D}$, as a resonance in the $D\bar{D}$ invariant mass spectrum~\cite{Belle:2005rte, BaBar:2010jfn}. Angular distribution analyses indicate that the $J^{PC}$ of this state favors $2^{++}$, consistent with the $\chi_{c2}(2P)$ assignment. In the amplitude analysis of $B\to D^+D^- K$ events by the LHCb Collaboration, the quantum numbers of this state were determined to be $2^{++}$, reinforcing its assignment as $\chi_{c2}(2P)$. 

\end{itemize}

\subsection{$Y$ states}\label{sec6:subsec63}

\begin{figure}[htb]
	\centering
	\includegraphics[width=2.0 in, height=2.5 in, angle = -90]{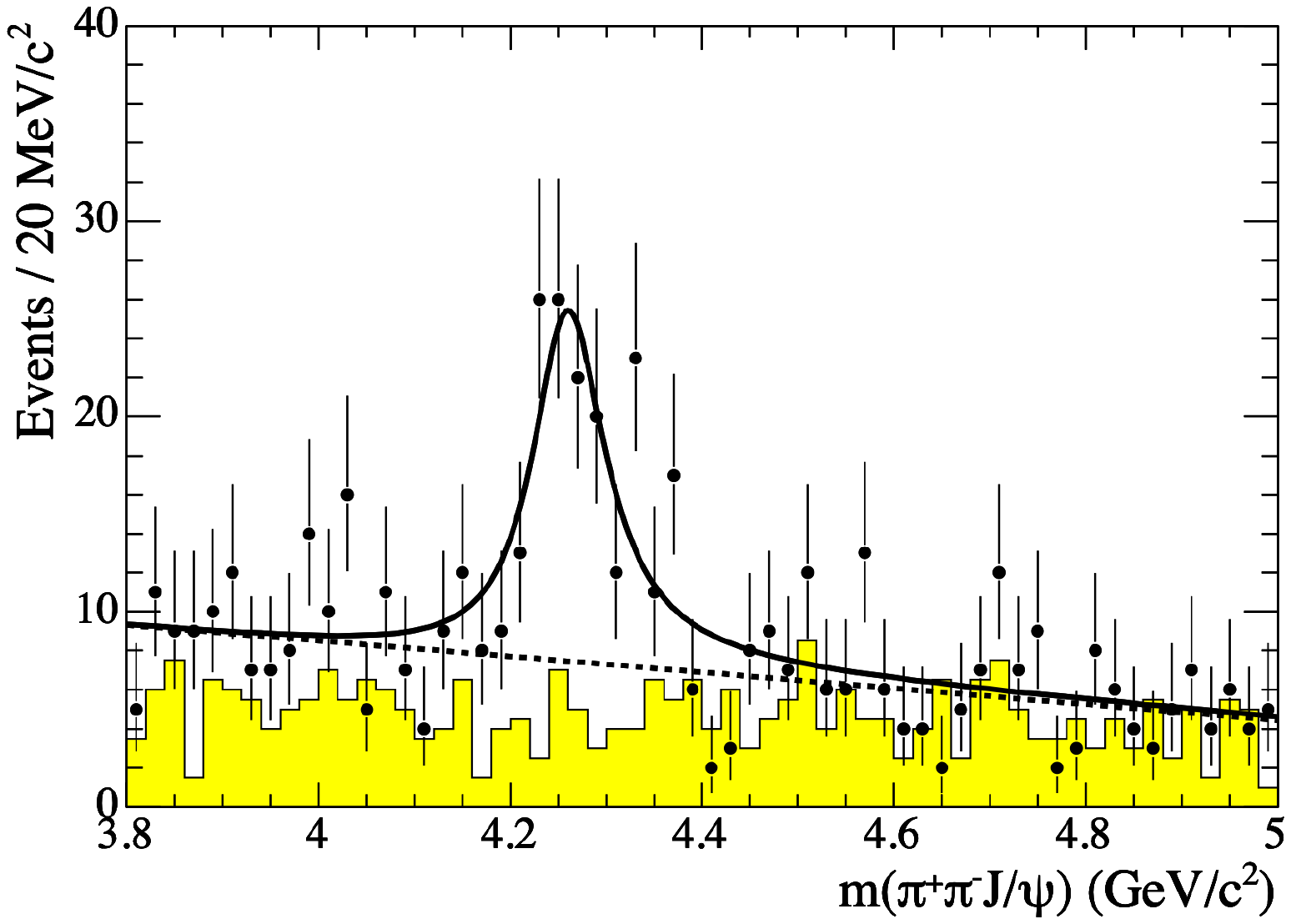}
	\caption{Invariant mass distribution of $J/\psi \pi^+\pi^-$ from the ISR process $e^+e^-\to\gamma_{\rm ISR}J/\psi \pi^+\pi^-$~\cite{BaBar:2005hhc}. The shaded histogram represents background events from non-$J/\psi$ process.}
	\label{sec6:fig2}
\end{figure}

The symbol $Y$ is used to denote a vector state with $J^{PC}=1^{--}$. According to the PDG naming convention, they are formally labeled as $\psi(\mathrm{mass})$. The first $Y$ state, $Y(4260)$, was discovered by the BaBar Collaboration in the $J/\psi\pi^+\pi^-$ invariant mass spectrum via the ISR process~\cite{BaBar:2005hhc}
\begin{align}
 e^+e^-\to \gamma_{\rm ISR} + J/\psi\pi^+\pi^-,\, J/\psi \to e^+e^-, \, \mu^+\mu^-.
\end{align}
The corresponding invariant mass distribution is shown in Fig.~\ref{sec6:fig2}, where a clear resonance peak appears around $4.26~\mathrm{GeV}$ above the background. The analysis used data samples collected near the $\Upsilon(4S)$ resonance, with the ISR photon effectively reducing the c.m. energy at which the $e^+e^-$ annihilation occurs. This observation was subsequently confirmed by both the CLEO and Belle Collaborations using the same decay channel. The quantum numbers of the $Y$ states are unambiguously determined by the production mechanism. 

By replacing the $J/\psi$ with a $\psi(2S)$ in the final state, the BaBar Collaboration reported evidence for a different vector resonance, $Y(4320)$, in the $\psi(2S)\pi^+\pi^-$ invariant mass spectrum~\cite{BaBar:2006ait}. The structure was found to be a new vector state, $Y(4360)$, by the Belle Collaboration, which also reported the observation of a third vector resonance, $Y(4660)$, above $4.0~\mathrm{GeV}$~\cite{Belle:2007umv}. 

These $Y$ states exhibit several intriguing properties:
\begin{enumerate}%

\item Strong coupling to hidden-charm final states: Unlike conventional charmonium states above the open-charm threshold---such as $\psi(3770)$, which predominantly decay into open-charm final states containing a pair of charmed meson---the $Y$ states couple strongly to hidden-charm final states containing a charmonium. 
	
\item Absence in inclusive hadronic cross section: No distinct peaking structures corresponding to the $Y(4260)$, $Y(4360)$, or $Y(4660)$ have been observed in the inclusive hadronic cross section $e^+e^-\to\mathrm{hadrons}$.
		
\item Overpopulation of vector states: Potential models predict five vector charmonium states between the $\psi(1D)$ and $4.7~\mathrm{GeV}$---namely, the $3S$, $2D$, $4S$, $3D$, and $5S$ states. However, experimental observations have revealed more than six such states: $Y(4260)$, $Y(4360)$, and $Y(4660)$ in exclusive processes, and $\psi(4040)$, $\psi(4160)$, and $\psi(4415)$ in inclusive hadronic cross section measurement~\cite{BES:2007zwq}. 

\item Process-dependent line shapes: The cross section line shapes vary significantly across different exclusive processes. Not all well-established $Y$ states appear in all dominant decay channels, highlighting strong process dependence and suggesting an important role of coupled-channel effects.

\end{enumerate}
 
\subsubsection{Fine structure of $Y(4260)$ and multiple decay modes of $Y(4230)$}\label{sec6:subsubsec631}

A major advantage of $e^+e^-$ colliders is the ability to perform energy scan, allowing the study of $Y$ states through cross section line shapes. This approach, reminiscent of the original discovery of  $J/\psi$ at SLAC, has been applied at the BESIII experiment. The BESIII experiment has collected $e^+e^-$ collision data across a c.m. energy range from $1.84$ to $4.95~\mathrm{GeV}$, with a fine scan between $3.8$ and $4.95~\mathrm{GeV}$. These fine scan data sets include 199 energy points with a total integrated luminosity of 26~fb${}^{-1}$, enabling cross section measurements for a wide array of exclusive processes, $e^+e^-\to X_i$, covering hidden-charm, open-charm, and light hadron final states. 

In 2017, the BESIII Collaboration measured the $e^+e^- \to J/\psi\pi^+\pi^-$ cross sections and found that the previously observed $Y(4260)$ actually consists of two resonant structures: one around $4.23~\mathrm{GeV}$ with a narrower width than observed in ISR-based measurements, and the other around $4.32~\mathrm{GeV}$~\cite{BESIII:2016bnd}. The lower-mass state was renamed $Y(4230)$ ($\psi(4230)$ in PDG). The higher-mass state exhibits parameters compatible with those of the $Y(4360)$ observed in the $\psi(2S)\pi^+\pi^-$ channel from the ISR process.
 
At almost the same time, two vector states were observed in the cross section of $e^+e^-\to h_c(1P)\pi^+\pi^-$. The lower-mass state is consistent with $Y(4230)$, indicating its decay into multiple final states. The higher-mass resonance, located near $4.39~\mathrm{GeV}$, was referred to as $Y(4390)$~\cite{BESIII:2016adj}. A similar state with mass around $4.22~\mathrm{GeV}$ was previously observed in $e^+e^-\to \chi_{c0}(1P)\omega$ process. 
A precise cross section measurement of $e^+e^-\to\psi(2S)\pi^+\pi^-$ using these fine scan data sets confirmed the presence of $Y(4360)$ and $Y(4660)$, 
and for the first time observed $Y(4230)$ in this channel.

To date, the $Y(4230)$ has been studied in over 70 exclusive final states and has been seen in 11 of them. These include both hidden-charm and open-charm final states: $\eta_{c}(1S)\pi^+\pi^-\pi^0$, $J/\psi\pi\pi$, $J/\psi\eta$, $J/\psi K\bar{K}$, $h_c(1P)\pi\pi$, $h_c(1P)\eta$, $\chi_{c0}(1P)\omega$, $\psi(2S)\pi\pi$, $D^0 D^{*-}\pi^+$, $D^{*}\bar{D}^{*}\pi$, and $\chi_{c1}(3872)\gamma$. So far, its presence in open-charm final states has only been observed in three-body decays, and it has not yet been seen in any purely light hadron final states~\cite{ParticleDataGroup:2024cfk}. 

\subsubsection{Parametrization of the cross section line shape and coupled-channel effects}\label{sec6:subsubsec632}

\begin{figure}[htb]
    \centering
    \includegraphics[width=6.0 in, height=2.5 in ]{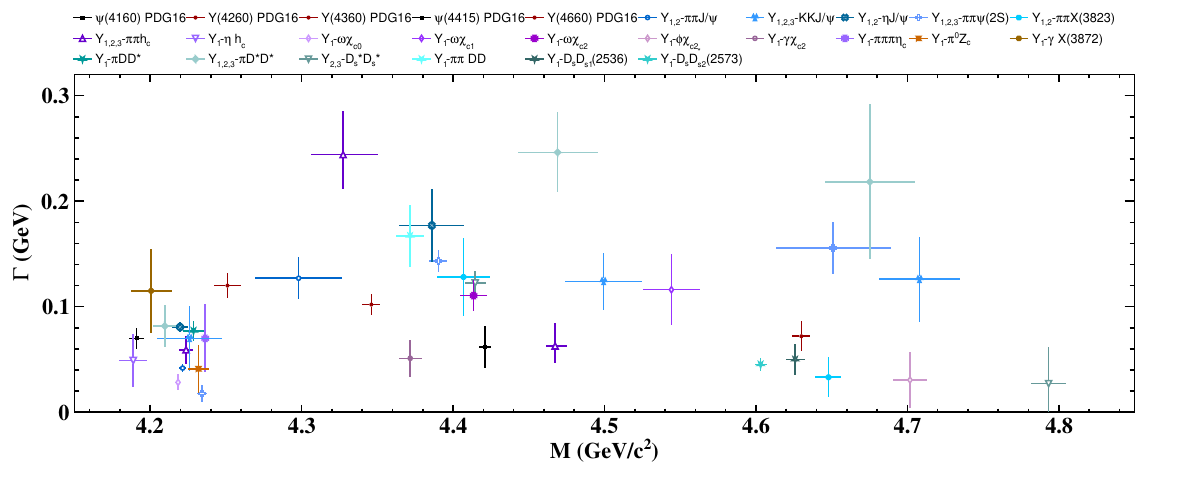}	
    \caption{The parameters of the vector states obtained from single-channel analyses.}
	\label{sec6:fig3}
\end{figure}

The resonance parameters of the $Y$ states are typically determined by fitting the cross section of the exclusive process using a model that includes Breit-Wigner functions for the resonant structures and a power-law term $1/s^{n}$ for the continuum contribution. The parametrization takes the form
\begin{align}
      \sigma(s)=\left |\frac{c}{s^{n}}+\sum_{k}BW_{k}(s)e^{i\phi_k}\right |^2,
\end{align}
where $c$ and $n$ are free parameters determined by the fit, and $\phi_k$ is the relative phase between the continuum and resonant amplitudes. The number of resonant structures included in the fit is guided by a hypothesis test, with additional resonances incorporated only if their statistical significance exceeds $5\sigma$. Interference effects are included when necessary. The extracted mass and width of the $Y$ states from each exclusive process are summarized in Fig.~\ref{sec6:fig3}.

A large number of $Y$ states have been observed through the ISR processes by the Belle and BaBar Collaborations, and using direct $e^+e^-$ annihilation with fine scan data sets by the BESIII Collaboration. However, none of these states have yet been seen via other production mechanisms. In addition to the well-known $Y(4260/4230)$, $Y(4360)$, and $Y(4660)$ states, several new structures have been reported:
\begin{itemize}
    \item $Y(4500)$ and $Y(4710)$ in the $e^+e^-\to J/\psi K\bar{K}$ process,
    \item $Y(4630)$ in $D_s^+ D_{s1}(2536)^-$ and $D_s^+ D_{s2}^{*}(2573)^-$ process, and 
    \item $Y(4790)$ in $D_s^{*+} D_s^{*-}$ process. 
\end{itemize}
These states decay into final states containing $c\bar{c}s\bar{s}$ quark content, suggesting a possible underlying connection. 

In the vicinity of the $Y(4230)$, resonance parameters extracted from different processes are relatively consistent. However, at higher c.m. energies, the parameters of these $Y$ states vary significantly between channels. This discrepancy is likely due to the limitations of the current cross-section-fit strategy, which typically neglects coupled-channel effects that could play an important role~\cite{Eichten:1979ms}. A more rigorous coupled-channel analysis, incorporating precise measurements of two-body (and possibly three-body and four-body) open-charm cross sections as well, would provide a more reliable and comprehensive understanding of the $Y$ spectrum---though such an analysis poses substantial theoretical and experimental challenges. 

\subsubsection{Vector states in bottomonium region}\label{sec6:subsubsec633}

In the bottomonium sector, three vector states have been observed above the well-known $\Upsilon(4S)$ state: $\Upsilon(10750)$, $\Upsilon(10860)$, and $\Upsilon(11020)$. The $\Upsilon(10750)$ was observed in the cross sections of the processes $e^+e^-\to\Upsilon(nS)\pi^+\pi^-$ ($n=1,2,3$)~\cite{Belle:2019cbt}, and was confirmed through a global $K$-matrix fit to both exclusive and inclusive $e^+e^-\to b\bar{b}$ cross sections~\cite{Husken:2022yik}. An enhancement near $\Upsilon(10750)$ was also found in $e^+e^-\to\chi_{b1,2}(1P)\omega$ process~\cite{Belle-II:2022xdi}. 

The $\Upsilon(10860)$ and $\Upsilon(11020)$ were initially identified as the $5S$ and $6S$ bottomonium states, respectively. However, several observations challenge this interpretation: notably, the high transition rates of $\Upsilon(5,6S)\to \Upsilon(nS)\pi^+\pi^-$, which are two orders of magnitude larger than expected for pure bottomonium states, and the abnormal rates of $\eta$ transition to lower mass $\Upsilon$ states. These anomalies cast doubt on their assignment as pure $b\bar{b}$ bound states. At present, experimental information about these states remains limited.

\subsection{$Z_{c}$ and $Z_{b}$ states}\label{sec6:subsec64}

While the $X$ and $Y$ states resemble traditional quarkonium states in that they are electrically neutral and carry traditional quantum numbers, the $Z_c$ and $Z_b$ states are clearly exotic: they are charged and decay into final states containing traditional heavy quarkonia. These characteristics imply that such states must contain a $Q\bar{Q}$ pair, since otherwise their decays would be suppressed by the OZI rule. Consequently, their minimal quark composition should be four, $c\bar{c}q\bar{q}$ for $Z_c$ and $b\bar{b}q\bar{q}$ for $Z_b$. 
Table~\ref{sec6:tab2} summarizes the $Z_c$ and $Z_b$ states observed in experiments. 

\begin{table}[htb]
	\TBL{\caption{Basic properties of $Z_c$ and $Z_b$ states. ``Production" and ``Decay" refer to all the reported production mechanisms and decay channels; others are the same as in Table~\ref{sec4:tab1}.}
	\label{sec6:tab2}}
	{\begin{tabular*}{\textwidth}{@{\extracolsep{\fill}}@{}lllllll@{}}
			\toprule
			\multicolumn{1}{@{}l}{\TCH{Name}} &
			\multicolumn{1}{l}{\TCH{M (MeV)}} &
			\multicolumn{1}{l}{\TCH{$\Gamma$ (MeV)}} &
			\multicolumn{1}{l}{\TCH{$I^{G}J^{PC}$}} &
			\multicolumn{1}{l}{\TCH{Production}} &
			\multicolumn{1}{l}{\TCH{Decay}} &
			\multicolumn{1}{l}{\TCH{Experiment}}\\
			\toprule
			$Z_{c}(3900)$	& $3887.1\pm2.6$		& $28.4\pm2.6$ 	& $1^+(1^{+-})$		& $e^{+}e^{-}\to\pi^{\pm} Z_{c}(3900)^{\mp}$	& $J/\psi\pi^{\mp}$	& BESIII, Belle  \\
						&                			&            			&          			& 									& $(D\bar{D})^{\mp}$ 	& BESIII \\
						&                			&            			&          			& $e^{+}e^{-}\to\pi^{0} Z_{c}(3900)^{0}$		& $J/\psi\pi^{0}$		& BESIII \\
                       				&                			&           			&          			& 									& $(D\bar{D})^{0}$ 		& BESIII \\
						&               			&           			&          			& 									& $\eta_c(1S)\rho$ 		& BESIII \\
                         				&           				&             			&          			& $H_{b}\to Z_c(3900)X$					& $J/\psi\pi$ 			&   D\O \\ 
			$Z_{c}(4020)$ 	& $4024.1\pm1.9$ 		& $13\pm5$     		& $1^+(?^{?-})$ 	& $e^{+}e^{-}\to\pi^{\pm} Z_c(4020)^{\mp}$ 	& $h_c(1P)\pi^{\mp}$ 	& BESIII  \\
                      				&                			&              			&          			& 									& $(D^{*}\bar{D}^{*})^{\mp}$ & BESIII \\
						&                			&              			&          			& $e^{+}e^{-}\to\pi^{0} Z_c(4020)^{0}$		& $h_c(1P)\pi^{0}$ 		& BESIII \\
						&                			&              			&          			& 									& $(D^{*}\bar{D}^{*})^{0}$ & BESIII \\
			$Z_c(4050)\dag$& $4051^{+24}_{-40}$ 	& $82^{+50}_{-28}$	& $1^{-}(?^{?+})$	& $B\to K Z_c(4050)$					& $\chi_{c1}(1P)\pi$ 		& Belle\\
			$Z_c(4055)\dag$& $4054\pm 3.2$ 	     	& $45\pm 13$		& $1^{+}(?^{?-})$ 	& $e^{+}e^{-}\to\pi^{\pm} Z_{c}(4055)^{\mp}$	& $\psi(2S)\pi^{\mp}$ 	& Belle\\
			$Z_c(4100)\dag$& $4096\pm28$   		& $150^{+80}_{-70}$	& $1^{-}(?^{?+})$	& $B\to K Z_c(4100)$					& $\eta_c(1S)\pi$	& LHCb \\
			$Z_c(4200)\dag$& $4196^{+35}_{-32}$ 	& $370^{+100}_{-150}$ & $1^{+}(1^{+-})$ 	& $B\to KZ_c(4200)$						& $J/\psi\pi$ 		& Belle\\
			$R_{c0}(4240)\dag$& $4239^{+50}_{-21}$ & $220^{+120}_{-90}$ & $1^+(0^{--})$	& $B\to K R_{c0}(4240)$					& $\psi(2S)\pi$ 		& LHCb\\
			$Z_c(4250)\dag$& $4250^{+190}_{-50}$ 	& $180^{+320}_{-70}$ & $1^-(?^{?+})$ 	& $B\to KZ_c(4250)$						& $\chi_{c1}(1P)\pi$  & Belle\\
			$Z_c(4430)$	& $4478^{+15}_{-18}$ 	& $181\pm31$ 		& $1^+(1^{+-})$ 	& $B\to KZ_c(4430)$						& $\psi(2S)\pi$ 		& Belle, LHCb \\
                     				&                  			&             			&          			&									& $J/\psi\pi$ 		& Belle\\
			
			\midrule
			$Z_{cs}(3985)\dag$  & $3982.5^{+2.8}_{-3.4}$ & $12.8^{+6.1}_{-5.3}$ & $1/2(?^{?})$ 	& $e^{+}e^{-}\to K Z_{cs}(3985)$	& $D_{s}^{-}D^{*0}+D_{s}^{*-}D^{0}$  & BESIII\\
                  	$Z_{cs}(4000)\dag$  & $4003^{+7}_{-15}$    &  $131\pm30$  		& $1/2(1^{+})$      	& $B\to\phi Z_{cs}(4000)$			& $J/\psi K$ 		& LHCb\\
			$Z_{cs}(4220)\dag$  & $4220^{+50}_{-40}$   &  $230^{+110}_{-90}$   & $1/2(1^{+})$       & $B\to\phi Z_{cs}(4220)$			& $J/\psi K$ 		& LHCb\\
			\midrule
			$Z_b(10610)$		 & $10607.2\pm2.0$    &  $18.4\pm2.4$  		& $1^+ (1^{+-})$      	& $e^+e^-\to \pi^{\pm} Z_b(10610)^{\mp}$	& $\Upsilon(1,2,3S)\pi^{\mp}$, $h_b(1,2P)\pi^{\mp}$ 		& Belle\\
							&                  			&             			&          		&  								& $(B\bar{B}^*)^{\mp}$ 		& Belle\\
			$Z_b(10650)$		 & $10652.2\pm1.5$    &  $11.5\pm2.2$  		& $1^+ (1^{+-})$      & $e^+e^-\to\pi^{\pm} Z_b(10610)^{\mp}$		& $\Upsilon(1,2,3S)\pi^{\mp}$, $h_b(1,2P)\pi^{\mp}$ 		& Belle\\
							&                  			&             			&          		&  								& $(B^{*}\bar{B}^*)^{\mp}$ 		& Belle\\
			\botrule
\end{tabular*}}{%
	}%
\end{table}

\subsubsection{$Z_c$ states}\label{sec6:subsubsec641}

\begin{figure}[htb]
	\centering
	\includegraphics[width=3.5 in, height=5.2 in, angle=-90 ]{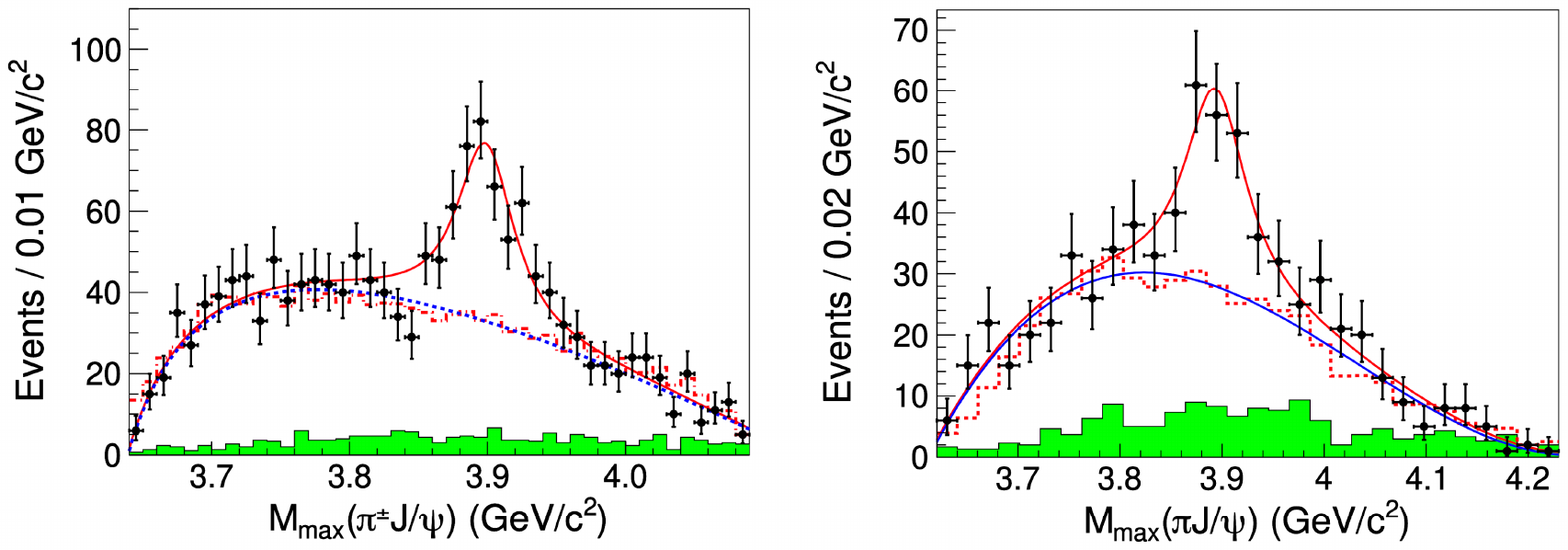}
	\caption{The invariant mass distributions of $J/\psi\pi$ from BESIII (left)~\cite{BESIII:2013ris} and Belle (right)~\cite{ Belle:2013yex} measurements. The shaded histograms represent background events from non-$J/\psi$ process. }
	\label{sec6:fig4}
\end{figure}

In 2013, a charged charmonium-like state, $Z_c(3900)$ ($T_{c\bar{c}1}(3900)$ in PDG), was simultaneously observed by the BESIII Collaboration in the reaction~\cite{BESIII:2013ris}
\begin{align}
  e^+e^- \to Y(4260) \to J/\psi \pi^{\pm}  + \pi^{\mp},
\end{align}
and by the Belle Collaboration via the ISR process~\cite{ Belle:2013yex} 
\begin{align}
  e^+e^- \to \gamma_{\rm ISR} + Y(4260) \to \gamma_{\rm ISR} + J/\psi \pi^{\pm}  + \pi^{\mp}.
\end{align}
The $Z_c(3900)$ appears as a clear peak in the $J/\psi \pi^{\pm}$ invariant mass spectrum above background events. Projections of $M_{\rm max}(J/\psi \pi^{\pm})$, which is the higher mass combination of the $J/\psi \pi^+$ and $J/\psi \pi^-$ in the event, from both experiments are shown in Fig.~\ref{sec6:fig4}. 

The mass and width of $Z_c(3900)$ measured by the two experiments are in good agreement:
\begin{align*} 
 &M=3899.0\pm3.6\pm4.9~\mathrm{MeV}, ~\Gamma=46\pm10\pm20~\mathrm{MeV}~\mathrm{(BESIII)},\\
 &M=3894.5\pm6.6\pm4.5~\mathrm{MeV}, ~\Gamma=63\pm24\pm26~\mathrm{MeV}~\mathrm{(Belle)}. 
\end{align*}
Shortly thereafter, this state was confirmed using CLEO-c data taken at $\sqrt{s}=4.17~\mathrm{GeV}$. Evidence for its neutral partner was also reported using the same decay channel, replacing $\pi^+\pi^-$ with $\pi^0\pi^0$~\cite{Xiao:2013iha}. 

The mass of $Z_c(3900)$ is close to the $D\bar{D}^{*}$ mass threshold, and as expected, this state was also observed in the open-charm process $e^+e^-\to \pi^{\pm} (D\bar{D}^{*})^{\mp}$~\cite{BESIII:2013qmu}, in the $D\bar{D}^{*}$ mass spectrum close to the threshold. Its neutral partner was also observed with significance above $5\sigma$ by the BESIII Collaboration in both hidden-charm and open-charm final states, confirming isospin symmetry. An amplitude analysis by the BESIII Collaboration determined its quantum numbers, favoring $J^{P}=1^{+}$ over other quantum numbers with a significance exceeding 7$\sigma$~\cite{BESIII:2017bua}. 

Also in 2013, the BESIII Collaboration discovered a second $Z_c$ state with higher mass, $Z_c(4020)$ ($T_{c\bar{c}}(4020)$ in PDG), in the process:
\begin{align}
    e^+e^-\to h_c(1P)\pi^+\pi^-,
\end{align}
where a distinct resonance peak was observed above the background contribution~\cite{BESIII:2013ouc}. This observation was made using data sets collected near the $Y(4230)$ and $Y(4360)$ peaks. The mass of the $Z_c(4020)$ is close to the $D^{*}\bar{D}^{*}$ threshold, again this state was also observed in $D^{*}\bar{D}^{*}$ final state by the BESIII Collaboration~\cite{BESIII:2013mhi}. 

The discovery of states containing at least four quarks marked a major breakthrough in particle physics. The observation of such ``Four-Quark Matter" was selected by the American Physical Society as one of the ``Highlights of the Year" in 2013.

Interestingly, the first reported candidate for a four-quark hidden charm state was $Z_c(4430)$, observed by the Belle Collaboration in the $B$ meson decay process, $B\to \psi(2S)\pi K$~\cite{Belle:2007hrb}. Although initial confirmation attempts failed, the state was rediscovered by the LHCb Collaboration in 2014 using the same channel.

The minimal quark content of the three $Z_c$ states---$Z_c(3900)$, $Z_c(34020)$, and $Z_c(4430)$, is $c\bar{c}u\bar{d}$ and $c\bar{c}d\bar{u}$, depending on the charge. Assuming SU(3) flavor symmetry, strange counterparts of these states, known as $Z_{cs}$, should exist. 

Such states were observed by the BESIII Collaboration in the reaction
\begin{align}
  e^+e^-\to  K^+(D_s^-D^{*0}+D_s^{*-}D^0) + c.c.
\end{align} 
in the $(D_s^-D^{*0}+D_s^{*-}D^0)+c.c.$ invariant mass spectrum~\cite{BESIII:2020qkh} and by the LHCb Collaboration in the decay
\begin{align} 
  B\to J/\psi K + \phi
\end{align} 
in the $J/\psi K$ invariant mass spectrum~\cite{LHCb:2021uow}.
These states were identified as $Z_{cs}(3985)$ and $Z_{cs}(4000)$, respectively. While their masses are compatible, their widths differ significantly---approximately 10~MeV for $Z_{cs}(3985)$ and 130~MeV for $Z_{cs}(4000)$. PDG currently lists them under a single entry, $T_{c\bar{cs}1}(4000)$, although direct confirmation of $Z_{cs}(4000)$ in open-charm final state and $Z_{cs}(3985)$ in hidden-charm final state remains pending. The quantum numbers of $Z_{cs}(4000)$ were determined by LHCb to be $J^{P}=1^{+}$.

\subsubsection{$Z_{b}$ states}\label{sec6:subsubsec642}

In the bottomonium sector, two charged bottomonium-like states, $Z_b(10610)$ and $Z_b(10650)$, were observed by the Belle Collaboration in the processes
\begin{align}
    e^+e^-\to \Upsilon(nS)\pi^+\pi^-~(n=1,2,3),
\end{align}
and 
\begin{align}
    e^+e^- \to h_b(nP)\pi^+\pi^-,~(n=1,2),
\end{align}
using data collected near the $\Upsilon(10860)$ peak~\cite{Belle:2011aa}. The quantum numbers $J^{P}=1^{+}$ are strongly favored for both states. 

These two states are located near the open-beauty thresholds: $Z_b(10610)$ close to $B\bar{B}^*$ threshold whereas $Z_b(10650)$ near $B^{*}\bar{B}^{*}$ threshold. Both states have been observed to decay into open-beauty final states.  

\subsection{$T_{cc}$ states}\label{sec6:subsec65}

The minimal quark content of the $Z_c$ and $Z_b$ states is $c\bar{c}q\bar{q}$ and $b\bar{b}q\bar{q}$, respectively, where there are a heavy quark and a heavy antiquark of the same flavor in them. The existence of four-quark states composed of two heavy quarks and two light antiquarks, denoted as $T_{cc}$, has been discussed in the literature. For the ground $cc\bar{u}\bar{d}$ state with quantum numbers $J^P=1^+$ and isospin $I=0$, the mass is predicted to lie close to the $D^{*+}D^0$ threshold. The mass difference, $\delta m \equiv m_{T_{cc}^+}-(m_{D^{*+}}+m_{D^0})$, is expected to be within the range of $(-300,~+300)~\mathrm{MeV}$, according to theoretical studies referenced in Ref.~\cite{LHCb:2021vvq}. 

\begin{figure}[htb]
	\centering
	\includegraphics[width=2.5 in, height=2.0 in]{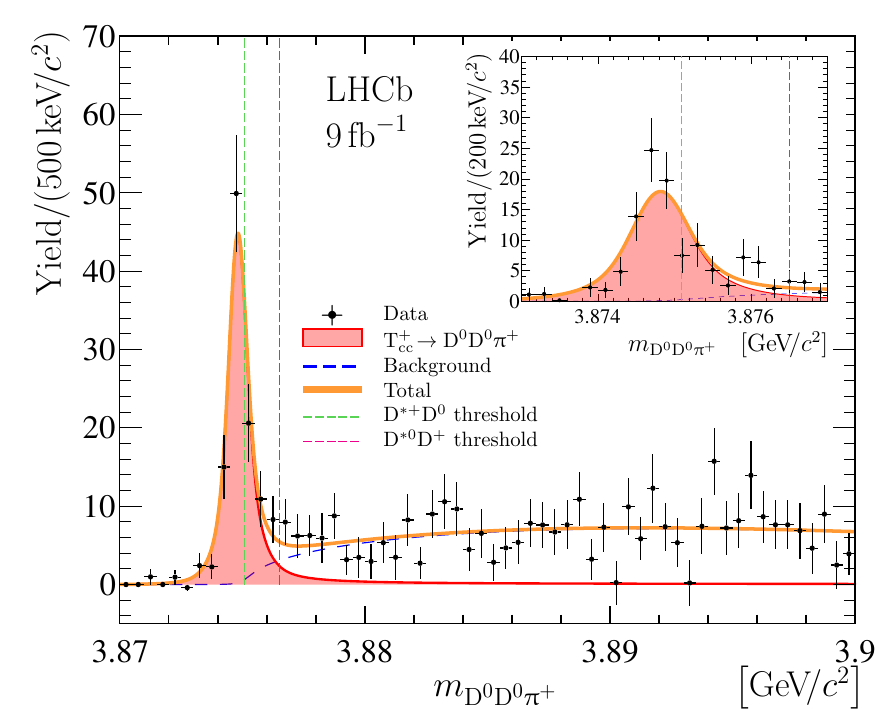}
	\caption{Invariant mass distribution of $D^0D^0\pi^+$ from $pp$ prompt production~\cite{LHCb:2021vvq}.}
	\label{sec6:fig5}
\end{figure}

In 2022, the LHCb Collaboration discovered a state consistent with the predicted $T_{cc}$ in the reaction 
\begin{align}
    pp\to T_{cc}^{+} + X \to D^{0} D^{0}\pi^+,
\end{align}
where a narrow peak was observed in the invariant mass spectrum of $D^0D^0\pi^+$~\cite{LHCb:2021vvq}, as shown in Fig.~\ref{sec6:fig5}. The signal lies just below the $D^{*+}D^0$ threshold, and the mass difference was measured to be $\delta m=-273\pm61\pm5^{+11}_{-14}~\mathrm{keV}$ with a decay width of $\Gamma=410\pm165\pm43^{+18}_{-38}~\mathrm{keV}$. These values were obtained by fitting the invariant mass spectrum with a Breit-Wigner function incorporating threshold effects for the signal component. The third uncertainty is related to the assumed quantum numbers $J^P=1^+$. 
While taking into account the fact that the resonance is close to the $D^{*}D$ threshold, and employing a unitarised Breit-Wigner function constructed under the assumptions that the state has quantum numbers $I (J^{P})=0 (1^{+})$ and couples strongly to the $D^{*}D$ channel, the pole parameters are determined to be $\delta m_{\rm pole}=-360\pm40\pm5^{+4}_{-0}~\mathrm{keV}$ and $\Gamma_{\rm pole}=48\pm2^{+0}_{-14}~\mathrm{keV}$~\cite{LHCb:2021auc}. 
This is the narrowest exotic state observed to date. In contrast, other $XYZ$ states typically exhibit widths from several tens $\mathrm{MeV}$ to a few hundred $\mathrm{MeV}$. 

\subsection{$P_c$ states}\label{sec6:subsec66}

The $P_c$ pentaquark states with the minimal quark content of $c\bar{c}uud$ were discovered by the LHCb Collaboration in 2015~\cite{LHCb:2015yax} and 2019~\cite{LHCb:2019kea} in the decay process
\begin{align}\label{sec6:eq1}
   \Lambda_b^0 \to J/\psi p K^-,
\end{align}
where the dominant contribution was expected to arise from intermediate excited baryons
\begin{align}\label{sec6:eq2}
   \Lambda_b^0 \to J/\psi + \Lambda^*, ~\Lambda^* \to K p,
\end{align}
with $\Lambda^*$ stands for any resonant excitation of $\Lambda$ baryon. However, it is found that using processes shown in Eq.~\ref{sec6:eq2} alone can not describe data well, and three narrow resonances, $P_c(4312)$, $P_c(4440)$, and $P_c(4457)$ were observed in the $J/\psi p$ invariant mass spectrum. 

Pentaquark state with a strange quark ($c\bar{c}uds$), referred to as $P_{cs}$, was observed in the $B$-meson decays through the $J/\psi \Lambda$ invariant mass spectrum, further enriching the spectrum of exotic baryonic states.


To date, all known $P_c$ states were observed by the LHCb Collaboration in $b$-hadron decays and in a single decay mode. It is essential to explore additional production mechanisms and decay channels to gain a more comprehensive understanding of these states. Table~\ref{sec6:tab3} summarizes the $P_c$ states observed from experiments. 

\begin{table}[htb]
	\TBL{\caption{Basic properties of $P_c$. ``Production" and ``Decay" refer to all the reported production mechanisms and decay channels, others are the same as in Table~\ref{sec4:tab1}.}
	\label{sec6:tab3}}
	{\begin{tabular*}{\textwidth}{@{\extracolsep{\fill}}@{}llllll@{}}
			\toprule
			\multicolumn{1}{@{}l}{\TCH{Name}} &
			\multicolumn{1}{l}{\TCH{M (MeV)}} &
			\multicolumn{1}{l}{\TCH{$\Gamma$ (MeV)}} &
			\multicolumn{1}{l}{\TCH{$I^{G}J^{PC}$}} &
			\multicolumn{1}{l}{\TCH{Production}} &
			\multicolumn{1}{l}{\TCH{Decay}} \\
			\toprule
			$P_c(4312)\dag$	& $4311.9^{+7.0}_{-0.9}$	& $10\pm5$ 		& $1/2(?^{?})$		& 	$\Lambda_b\to K  P_c(4312)$		& $J/\psi p$	 \\
			$P_c(4440)\dag$	& $4440^{+4}_{-5}$		& $21^{+10}_{11}$ 	& $1/2(?^{?})$	& 	$\Lambda_b\to K  P_c(4440)$		& $J/\psi p$	 \\
			$P_c(4457)\dag$	& $4457.3^{+4.0}_{-1.8}$	& $6.4^{+6.0}_{2.8}$ 	& $1/2(?^{?})$	& 	$\Lambda_b\to K  P_c(4457)$		& $J/\psi p$	 \\
			\midrule
			$P_{cs}(4338)\dag$	& $4338.2\pm0.8$	& $7.0\pm1.8$ 		& $0(1/2^{-})$		& 	$B\to p  P_c(4338)$			& $J/\psi \Lambda$	 \\
			\botrule
\end{tabular*}}{%
	}%
\end{table}

\section{Conclusions}
\label{sec:conclusions}


This chapter has provided an overview of heavy quarkonia and the growing spectrum of newly observed hadrons containing two heavy quarks. The discovery of the $J/\psi$ in 1974 confirmed the existence of the charm quark, while the subsequent observation of $\Upsilon$ established the bottom quark. These systems have served as unique laboratories for studying both perturbative and nonperturbative aspects of QCD. 

The spectroscopy of heavy quarkonia, particularly below open-flavor thresholds, is well described by potential models incorporating Coulomb-like short-distance interactions and confining forces at long-distance. However, above the threshold, the spectrum becomes considerably more intricate. The discovery of a variety of new states---collectively referred to as the $XYZ$ states---has revolutionized the landscape of hadron physics. Well-known examples include $X(3872)$, $Y(4260/4230)$, $Z_{c}(3900)$, and $P_c(4312)$, along with many other candidates, exhibit properties incompatible with the traditional quark-antiquark meson or three-quark baryon picture, suggesting the existence of exotic hadrons with more complex internal configurations. 

These exotic states are widely interpreted as candidates for tetraquarks, pentaquarks, hadron molecules, hybrids, or other exotic configurations. Their production mechanisms (e.g., $e^+e^-$ annihilation, hadron colliders, $B$ decays) and decay patterns provide valuable clues, yet their internal composition remains unresolved. 

Despite significant progress, several major challenges persist:
\begin{itemize}
    \item What is the internal structure of exotic hadrons?\\
    Are they compact multiquark states, loosely bound hadronic molecules, or admixtures of multiple configurations? Could some be artifacts of kinematic effects? Disentangling these scenarios requires high-precision measurements of resonance parameters, quantum number assignments, decay widths, and production rates, supported by systematic and precise theoretical calculations. 
    
    \item Where are the missing conventional states? \\
    Potential models predict more conventional quarkonium states than currently observed, especially for excited states. Are these states obscured by decay dynamics or simply beyond the reach of current experiments?
    
    \item Where are the missing exotic states? \\
    Beyond the known $XYZ$ and $P_c$ states, many more states, either with different quark-gluon configurations or with the same quark-gluon configuration but different quantum numbers, may exist. Where are they? How are they produced, and how do they decay? Especially, why have so few states been observed in the bottomonium sector? 
    
    \item What is the role of coupled-channel effects? \\
    Many exotic states appear near meson-meson (meson-baryon) thresholds. How do such thresholds affect resonance properties (e.g., $Y$ states)? Global analyses that incorporate data from multiple decay modes and production channels are essential for understanding these effects.  
    
    \item What can exotic states reveal about QCD confinement? \\
    Can exotic hadrons reveal new aspects of the color-confining nature of QCD and the role of gluonic degrees of freedom in hadron structure? 
\end{itemize}

\vspace{1em}
Addressing these questions will require a multifaceted approach. High-statistics experimental programs at BESIII, Belle~II, and LHCb will continue to play a leading role, together with advanced data analysis approaches. In parallel, theoretical advances, including lattice QCD calculations and improved theoretical modeling, will be crucial for interpreting new data and refining our understanding. Ultimately, resolving these questions may lead to transformative insights into the dynamics of strong interaction.

\begin{ack}[Acknowledgments]%
This work is supported in part by National Natural Science Foundation of China (NSFC) under contract No.~12361141819, No.~12375070, National Key R\&D Program of China under Contract No.~2024YFA1610504, and the Shanghai Leading Talent Program of Eastern Talent Plan under Contract No. JLH5913002.
\end{ack}


\bibliographystyle{Numbered-Style} 
\bibliography{reference}

\end{document}